\documentclass{bmcart}

\usepackage{amsthm,amsmath,rotating,multicol,multirow, booktabs,url,algorithm2e,mathtools,threeparttable,graphicx,subcaption}
\DeclarePairedDelimiter\ceil{\lceil}{\rceil}
\usepackage[utf8]{inputenc} 

\startlocaldefs
\endlocaldefs

\begin{document}

\begin{frontmatter}

\begin{fmbox}
\dochead{Research}


\title{Analytical method for detecting outlier evaluators}


\author[
  addressref={aff1},                   
  email={yujiewu@g.harvard.edu}   
]{\inits{Y. W.}\fnm{Yujie} \snm{Wu}}
\author[
  addressref={aff3, aff5},
  email={scurhan@bwh.harvard.edu }
]{\inits{S. G. C.}\fnm{Sharon G.} \snm{Curhan}}
\author[
addressref={aff1,aff3},
email={stbar@channing.harvard.edu }
]{\inits{B. R.}\fnm{Bernard} \snm{Rosner}}
\author[
addressref={aff2, aff3, aff4, aff5},
email={gcurhan@bwh.harvard.edu }
]{\inits{G. G. C.}\fnm{Gary G.} \snm{Curhan}}
\author[
addressref={aff1, aff2, aff3},                   
corref={aff1},                       
email={stmow@channing.harvard.edu}   
]{\inits{M. W.}\fnm{Molin} \snm{Wang}}


\address[id=aff1]{
  \orgdiv{Department of Biostatistics},             
  \orgname{Harvard University},          
  \city{Boston},                              
  \cny{USA}                                    
}

\address[id=aff2]{%
  \orgdiv{Department of Epidemiology},
  \orgname{Harvard University},          
\city{Boston},                              
\cny{USA}   
}

\address[id=aff3]{%
	\orgdiv{Channing Division of Network Medicine},
	\orgname{Brigham and Women’s Hospital},
	\city{Boston},
	\cny{USA}
}
\address[id=aff4]{%
	\orgdiv{Renal Division, Department of Medicine,},
	\orgname{Brigham and Women’s Hospital},
	\city{Boston},
	\cny{USA}
}
\address[id=aff5]{%
	\orgname{Harvard Medical School},
	\city{Boston},
	\cny{USA}
}


\end{fmbox}


\begin{abstractbox}

\begin{abstract} 
	{\color{black}{
\parttitle{Background} 
Epidemiologic and medical studies often rely on evaluators to obtain measurements of exposures or outcomes for study participants, and valid estimates of associations depends on the quality of data. Even though statistical methods have been proposed to adjust for measurement errors, they often rely on unverifiable assumptions and could lead to biased estimates if those assumptions are violated. Therefore, methods for detecting potential `outlier' evaluators are needed to improve data quality during data collection stage. 

\parttitle{Methods} 
In this paper, we propose a two-stage algorithm to detect `outlier' evaluators whose evaluation results tend to be higher or lower than their counterparts. In the first stage, evaluators' effects are obtained by fitting a regression model. In the second stage, hypothesis tests are performed to detect `outlier' evaluators, where we consider both the power of each hypothesis test and the false discovery rate (FDR) among all tests.  We conduct an extensive simulation study to evaluate the proposed method, and illustrate the method by detecting potential `outlier' audiologists in the data collection stage for the Audiology Assessment Arm of the Conservation of Hearing Study, an epidemiologic study for examining risk factors of hearing loss in the Nurses' Health Study II.

\parttitle{Results}
Our simulation study shows that our method not only can detect true `outlier' evaluators, but also is less likely to falsely reject true `normal' evaluators.

\parttitle{Conclusions}
Our two-stage `outlier' detection algorithm is a flexible approach that can effectively detect `outlier' evaluators, and thus data quality can be improved during data collection stage.
}}
\end{abstract}




\end{abstractbox}
%

\end{frontmatter}



\section*{Introduction}
\noindent Many medical and epidemiological studies that investigate relationships between risk factors and disease outcomes rely on multiple evaluators (e.g. clinicians, technicians) to measure the exposures or outcomes of interest among study participants. For example, in large epidemiologic studies of hearing loss, pure-tone audiometry measurements are typically obtained by multiple audiologists or trained technicians in sound-treated booths \cite{cruickshanks1998prevalence, shargorodsky2010change, gopinath2010incidence}. Similarly, in large studies of vision, vision tests are often conducted by multiple evaluators in a clinic setting \cite{zhang2013association, klein2014relation}. Further, potential issues related to the collection of data by multiple evaluators may also extend to studies that rely on data collected by non-human testing methods, such as automated audiometers \cite{mccullough2019circulating}, to obtain test measurements. Obtaining precise estimates of the association between risk factors and disease outcomes not only depends on the statistical methods used, but also the quality of data itself. Although many analytical methods have been proposed to adjust for measurement errors arose from data collected with poor quality, those methods typically rely on unverifiable assumptions \cite{carroll2006measurement}, and pays a cost of the precision of estimates. Therefore, collecting data with better quality is preferred over using statistical methods to adjust for the biases induced by data of worse quality during statistical analysis stage. In this paper, we propose methods for quality control during data collection stage so that problems with the measurements of exposures or outcomes can be discovered and addressed promptly.

 Our work is motivated by the Conservation of Hearing Study (CHEARS), an investigation of risk factors for hearing loss among participants in the Nurses’ Health Studies II (NHS II), an ongoing cohort study consisting of 116,430 registered female nurses in the US, aged 25-42 years at enrollment in 1989 \cite{curhan2018adherence}. The CHEARS Audiology Assessment Arm (AAA) assessed the longitudinal change in the pure-tone air and bone conduction audiometric hearing thresholds (the sound intensity of a pure tone at which it is first perceived) measured in decibels in hearing level, or dB HL, across the full range of conventional frequencies (0.5-8 kHz) \cite{curhan2020prospective}. Baseline testing was conducted on 3,749 women whose self-reported hearing status was either `excellent', `very good' or had `a little hearing trouble', and resided within proximity of one of 19 CHEARS testing sites across the US \cite{curhan2020prospective}. The 3-year follow-up testing was completed on 3,136 participants (84\%). In order to obtain reliable hearing measurements, detecting potential `outlier' audiologists who tend to have higher or lower hearing test measurements than other audiologists is critical. Once an `outlier' audiologist is identified, devices used by this audiologist can be examined and an early intervention can be carried out during the data collection stage if necessary. Moreover, this outlier information may have important implications for the approach of data analysis.

 To the best of our knowledge, there are no existing statistical methods for detecting `outlier' evaluators. In this paper, we develop an innovative two-stage {\color{black}{algorithm}} for detecting `outlier' evaluators. In the first stage, rather than directly evaluating the observed measurements, we extract evaluators' effects on the measurements through regression analysis where the influences of other variables can be accounted for. In the second stage, we perform hypothesis tests to detect `outlier' evaluators based on the estimated coefficients and variances from the first-stage regression analysis.

 The paper is organized as follows. In Section `Methods', we present the two-stage {\color{black}{algorithm}} to detect `outlier' evaluators for scenarios when each study participant has either single or multiple measurements. In Section `Simulation', we perform a simulation study to investigate the performance of our two-stage {\color{black}{algorithm}}. Section `Application' presents a real data analysis to detect `outlier' audiologists in the CHEARS AAA. The section `Discussion' concludes the paper.

\section*{Methods}\label{section2}
\subsection*{First stage regression}\label{singe_outcome}
\noindent We first consider the scenario when each study participant only has one measurement to be obtained by an evaluator. Throughout the paper, we assumed that the exposure or test outcome of each study participant will be measured by only one evaluator, but one evaluator can measure multiple study participants. Let $i\in\{1,2,\ldots, N\}$ index the study participants; $j\in\{1,2,\ldots,M\}$ index the evaluators who measure the exposure or test outcome. Let $n_j$ denote the number of study participants who are evaluated by the $j$-th evaluator, such that $\sum_{j=1}^{M}n_j=N$. 

 To estimate the effects of evaluators on the measurements, in the first stage, we fit the following linear regression:
\begin{equation*}
	\text{E}(Y_i|\boldsymbol{X}_i, \text{T}_i^{(1)},\ldots, \text{T}_i^{(M)} )=\sum_{j=1}^M\beta_j\text{T}_i^{(j)} + \boldsymbol{\gamma}^T\boldsymbol{X}_i,
\end{equation*}
where $Y_i$ is the measurement for the $i$-th study participant, $\text{T}_i^{(j)}$ is an evaluator indicator which is 1 if the $i$-th study participant's exposure or outcome is evaluated by the $j$-th evaluator, and 0 otherwise, $\boldsymbol{X}_i$ is a $p$-dimensional vector containing potential confounders for the evaluator-$Y_i$ relationship and predictors of $Y_i$, and $\boldsymbol{\gamma}^T$ is the transpose of the $p$-dimensional coefficient vector $\boldsymbol{\gamma}$. Without further specification, all vectors are column vectors throughout this paper. Note that the first stage regression can go beyond linearity, where some nonlinear forms of $\boldsymbol{X}_i$ can be included for more accurate account of the effects of the covariates on the measurement. The regression coefficient $\beta_j$ represents the mean effect of evaluator $j$ on the measurement after adjusting for $\boldsymbol{X}$, and in the absence of `outlier' evaluators, $\beta_j, j=1,\ldots, M$, should be similar across different evaluators.

 In practice, there may be multiple measurements for all or part of study participants. Let $k\in\{1,2,\ldots,t_i\}$ index the measurements for the $i$-th study participant. For example, in the CHEARS AAA, study participants have both ears tested by audiologists, and therefore we have $t_i=2$ for each participant at each frequency.

 In the CHEARS AAA, the Pearson correlation coefficients between the hearing test outcomes of the left and right ear are over 0.7 regardless of frequencies. To take into account the correlation between multiple measurements while in the meantime being able to estimate the mean effect of evaluators on the measurements after controlling for potential confounders, we propose to apply the Generalized Estimating Equations (GEE) method in the first-stage regression analysis to estimate the effects of evaluators \cite{liang1986longitudinal, zeger1986longitudinal}. The model for the multiple correlated measurements can be written as:
\begin{equation*}
	\begin{split}
		E\left[{Y}_{i,k}|\boldsymbol{X}_{i}, \boldsymbol{Z}_{i,k},  T_{i}^{(1)},\ldots,T_{i}^{(M)}\right]&=\sum_{j=1}^M\beta_j\text{T}_i^{(j)}+\boldsymbol{\gamma}^T\boldsymbol{X}_i+\boldsymbol{\eta}^T\boldsymbol{Z}_{i,k},
	\end{split}
\end{equation*}
where  $\boldsymbol{Y}_i=[Y_{i,1},Y_{i,2},\ldots,Y_{i,t_i}]^T$, $\text{Cov}(\boldsymbol{Y}_i)=\Sigma_i$, with $\Sigma_i$ being the unknown $t_i\times t_i$ variance-covariance matrix of the measurements of the $i$-th study participant, and $\boldsymbol{Z}_{i,k}$ contains information that is specific to the $k$-th measurement of the $i$-th study participant. 

 The parameters $\boldsymbol{\theta}=[\boldsymbol{\gamma}^T, \boldsymbol{\beta}^T, \boldsymbol{\eta}^T]^T $, with $\boldsymbol{\beta}=[\beta_1,\ldots,\beta_M]^T$, can be estimated by solving the following estimating equation \cite{liang1986longitudinal, zeger1986longitudinal}:
\begin{equation*}
	\sum_{i=1}^{M}\boldsymbol{D}_i^T(\boldsymbol{\theta})\boldsymbol{V}_i^{-1}(\boldsymbol{\theta},\boldsymbol{\alpha})(\boldsymbol{Y}_i-\boldsymbol{\mu_i}(\boldsymbol{\theta}))=\boldsymbol{0},
\end{equation*}
where $\boldsymbol{\mu}_i=E\left[\boldsymbol{Y}_i|\boldsymbol{X}_{i}, \boldsymbol{Z}_i,  \text{T}_{i}^{(1)},\ldots,\text{T}_{i}^{(M)}\right]$, $\boldsymbol{D}_i=\frac{\partial}{\partial\boldsymbol{\theta}}\boldsymbol{\mu}_i(\boldsymbol{\theta})$, $\boldsymbol{V}_i(\boldsymbol{\theta},\boldsymbol{\alpha})$ is the working variance-covariance matrix, and $\boldsymbol{\alpha}$ contains parameters characterizing the correlation structure between multiple measurements. Some common working correlation structures for $k_1\ne k_2\in\{1,\ldots,t_i\}$ are independent, defined as $\text{Corr}(Y_{i,k_1}, Y_{i,k_2})=0$; exchangeable, defined as $\text{Corr}(Y_{i,k_1}, Y_{i,k_2})=\alpha$, and unstructured, defined as $\text{Corr}(Y_{i,k_1}, Y_{i,k_2})=\alpha_{k_1,k_2}$. The variance of $\widehat{\boldsymbol{\theta}}$, $\text{Var}(\widehat{\boldsymbol{\theta}})$, can be estimated based on the sandwich variance estimator \cite{liang1986longitudinal, zeger1986longitudinal}.

 The coefficients $\beta_1,\ldots,\beta_M$ reflect evaluators' mean effects on the measurements. An `outlier' evaluator will have a different coefficient than the remaining `normal' ones. Thus, in the second stage, we perform hypothesis tests to detect `outlier' evaluators based on $\widehat{\boldsymbol{\beta}}$ and $\widehat{\text{Var}}(\widehat{\boldsymbol{\beta}})$.




\subsection*{Hypothesis Testing}\label{hypo_mean}

 \noindent In the second stage, to detect `outlier' evaluators who have less accurate measurements than their counterparts, we perform hypothesis tests based on the coefficient estimates from the first stage regression. To get the effect of `normal' evaluators on the measurements, we take the average of the regression coefficients across all evaluators. Since $\beta_j, j=1,\ldots,M$ already represents the mean effect of the $j$-th evaluator on the measurements after controlling for study participants' characteristics, `outlier' evaluators can be defined as those whose effects on the measurements are statistically different from the mean effect averaged across all evaluators. Therefore, for a given evaluator $j$, the hypothesis can be formulated as:
\begin{equation*}
	H_{0, j}: \beta_j-\frac{1}{M}\sum_{q=1}^{M}\beta_q = 0,\quad j=1,2,\ldots,M.\,\,\, \text{     v.s.    } \,\,\, H_{1, j}: \beta_j-\frac{1}{M}\sum_{q=1}^{M}\beta_q\ne 0
\end{equation*}
which can be written as $H_{0, j}: \boldsymbol{L}^T_j\boldsymbol{\beta}=0\,\,\, \text{     v.s.    } \,\,\, H_{1, j}: \boldsymbol{L}^T_j\boldsymbol{\beta}\ne 0$, with 
\begin{equation*}
	\boldsymbol{L}_j=\begin{bmatrix}
		-\frac{1}{M} & -\frac{1}{M} &\ldots&\underbrace{\frac{M-1}{M}}_{\text{j-th location}}&-\frac{1}{M}&\ldots&-\frac{1}{M}
	\end{bmatrix}^T.
\end{equation*}

 Note that, $\beta_j-\frac{1}{M}\sum_{q=1}^{M}\beta_q$ can be interpreted as the difference between the mean measurement of the $j$-th evaluator and the average mean measurements over all evaluators adjusting for the characteristics of the study participants being evaluated. The test statistic of the Wald $\chi^2$ test under the null hypothesis $H_{0, j}$ is \cite{harrell2015regression}:
\begin{equation*}
	\left(\boldsymbol{L}_j^T\widehat{\boldsymbol{\beta}}\right)^T\left[  \boldsymbol{L}_j^T\widehat{\Sigma}\boldsymbol{L}_j \right]^{-1}\left(\boldsymbol{L}_j^T\widehat{\boldsymbol{\beta}}\right) \xrightarrow{D} \chi_1^2,
\end{equation*}
where $\widehat{\Sigma}$ is the estimated variance-covariance matrix of $\text{Var}(\widehat{\boldsymbol{\beta}})$.

 A more robust approach is to compute a truncated mean of the coefficients where potential `outliers' can be prevented from contaminating the average effect. Let $\beta_{(1)}, \beta_{(2)},\ldots,\beta_{(M)}$ be the ordered values of the regression coefficients. A $\delta\times 100 \%$ truncated mean can be calculated as follows \cite{wilcox2011introduction}:
$$\overline{\beta}_{\text{truncated}}=\frac{1}{M-2[M\cdot\delta]}\sum_{q=[M\cdot\delta]+1}^{M-[M\cdot\delta]}\beta_{(q)},$$
where $[x]$ denotes the integer part of $x$.

 The null hypothesis that the $j$-th evaluator is not an `outlier' is now to compare the regression coefficient of the $j$-th evaluator to the $\delta\times 100\%$ truncated mean:
\begin{equation*}
	H_{0, j}: \beta_{j}-\overline{\beta}_{\text{truncated}}= 0,\quad j=1, 2,\ldots, M.
\end{equation*}

 Let the set of the regression coefficients that are truncated be {\footnotesize{$\mathcal{A}=\{\beta_{(1)},\ldots,\beta_{([M\cdot\delta])}, \beta_{(M-[M\cdot\delta]+1)},\ldots, \beta_{(M)}\}$}}. It follows that the $l\text{-th}, l=1,\ldots,M,$ element of the contrast matrix for testing $H_{0, j}$ is
\begin{equation*}
	L_{\delta\times 100\%, jl}=\begin{cases}
		0, &\text{If } \beta_{l} \in \mathcal{A} \text{ and } l\ne j\\
		1, &\text{If } \beta_{l} \in \mathcal{A} \text{ and } l = j\\
		-\frac{1}{M-2[M\cdot\delta]},& \text{If } \beta_{l} \notin \mathcal{A} \text{ and } l\ne j\\
		1-\frac{1}{M-2[M\cdot\delta]}, &\text{If } \beta_{l} \notin \mathcal{A} \text{ and } l = j,\\
	\end{cases}
\end{equation*}
and the null hypothesis $H_{0, j}$ can be written as $H_{0, j}: \boldsymbol{L}^T_{\delta\times 100\%, j}\boldsymbol{\beta}=0$. Note that, we use $\boldsymbol{L}_{\delta\times100\%, j}, j=1,\ldots, M,$ to denote the contrast matrix, indicating that the hypothesis test is comparing each evaluator's regression coefficient with the $\delta\times100\%$ truncated mean of the regression coefficients of all evaluators.

 The Wald test statistic under the null hypothesis is:
\begin{equation*}
	\left(\boldsymbol{L}_{\delta\times 100\%, j}^T\widehat{\boldsymbol{\beta}}\right)^T\left[  \boldsymbol{L}_{\delta\times 100\%, j}^T\widehat{\Sigma}\boldsymbol{L}_{\delta\times 100\%, j} \right]^{-1}\left(\boldsymbol{L}^T_{\delta\times100\%, j}\widehat{\boldsymbol{\beta}}\right) \xrightarrow{D} \chi_1^2.
\end{equation*}

 Note that the contrast matrix $\boldsymbol{L}_{\delta\times 100\%, j}$ is not directly available since we need to know the ordering of the true coefficients $\beta_1,\ldots,\beta_M$ in advance. An approximation to $\boldsymbol{L}_{\delta\times 100\%, j}$ can be based on the estimated regression coefficients $\widehat{\beta}_1,\ldots,\widehat{\beta}_M$; that is,
\begin{equation*}
	L_{\delta\times 100\%, jl}\approx\begin{cases}
		0, &\text{If } \widehat\beta_{l} \in \mathcal{A}^\ast \text{ and } l\ne j\\
		1, &\text{If } \widehat\beta_{l} \in \mathcal{A}^\ast \text{ and } l = j\\
		-\frac{1}{M-2[M\cdot\delta]},& \text{If } \widehat\beta_{l} \notin \mathcal{A}^\ast \text{ and } l\ne j\\
		1-\frac{1}{M-2[M\cdot\delta]}, &\text{If } \widehat\beta_{l} \notin \mathcal{A}^\ast \text{ and } l = j,\\
	\end{cases}
\end{equation*}
where $\mathcal{A}^\ast=\{\widehat\beta_{(1)},\ldots,\widehat\beta_{([M\cdot\delta])}, \widehat\beta_{(M-[M\cdot\delta]+1)},\ldots, \widehat\beta_{(M)}\}$.

 Since our goal is to detect as many potential `outlier' evaluators as possible, we would like to achieve sufficient power when the evaluators are true `outliers'. Therefore, to complete the hypothesis testing procedure, different from the traditional approach where emphasis is placed upon controlling the type-I error $\alpha$ at an acceptable level, we also attach importance to ensuring an appropriate level of type-II error.

\subsection*{Type-I Error Determination}\label{power}

 \noindent Ideally, when performing hypothesis tests to detect potential `outlier' evaluators, there is sufficient power to reject the null hypotheses $H_{0,j}$ when a pre-specified alternative hypothesis $H_{1,j}$ is true. Denote the pre-specified alternative hypothesis as $H_{1,j}: \big|\boldsymbol{L}^T_j\boldsymbol{\beta}\big|= c$, where $c$ can be determined based on subject matter knowledge. For instance, in the CHEARS AAA, the `hearing threshold' for each individual ear is measured by the lowest sound intensity of a pure-tone signal presented individually to each ear, to which the listener reliably responds, and the pure-tone signal was measured in 5-dB steps \cite{curhan2020prospective}. As a result, hearing loss was defined as a greater than 5-dB HL increase in the pure-tone averages of testing frequencies at low-frequency (0.5, 1, 2 kHz), mid-frequency (3, 4 kHz), and high-frequency (6, 8 kHz) \cite{curhan2020prospective}. Therefore, it is important to identify audiologists who consistently gave 5-dB larger or smaller hearing test results than their counterparts after controlling for study participants' characteristics. Thus, a reasonable value for the alternative hypothesis for which we hope to have sufficient power to detect is $c=5$ for the CHEARS AAA. For presentational simplicity, we do not distinguish between $\boldsymbol{L}_j$ and $\boldsymbol{L}_{\delta\times 100\%, j}$ in this section, and we use $\boldsymbol{L}_j$ to denote the contrast matrix of both tests.

 In general, the power formula for the hypothesis test: $H_{0, j}: \boldsymbol{L}^T_j\boldsymbol{\beta}=0 \text{ v.s. } H_{1, j}:\Big|\boldsymbol{L}^T_j\boldsymbol{\beta} \Big|= c $ is:
\begin{equation*}
	\begin{split}
		\text{P}\left(   \left(\boldsymbol{L}^T_j\widehat{\boldsymbol{\beta}}\right)^T\left[  \boldsymbol{L}^T_j\widehat{\Sigma}\boldsymbol{L}_j \right]^{-1}\left(\boldsymbol{L}^T_j\widehat{\boldsymbol{\beta}}\right)>\chi_{1, 1-\alpha}^2 \Big{|} \Big|\boldsymbol{L}^T_j\boldsymbol{\beta}\Big|= c \right) 
		=\phi,
	\end{split}
\end{equation*}
where $\alpha$ is a two-sided type-I error rate, and $\phi$ is the power of the test.

 Under alternative hypothesis, test statistic $\left(\boldsymbol{L}^T_j\widehat{\boldsymbol{\beta}}\right)^T\left[  \boldsymbol{L}^T_j\widehat{\Sigma}\boldsymbol{L}_j \right]^{-1}\left(\boldsymbol{L}_j^T\widehat{\boldsymbol{\beta}}\right)$ follows a noncentral $\chi^2$ distribution with one degree of freedom and noncentral parameter $\lambda_j = \frac{c^2}{\boldsymbol{L}_j^T\widehat{\Sigma}\boldsymbol{L}_j}$ \cite{lehmann2006testing}; we denote this distribution as $\chi_1^2(\lambda_j)$. Let $F_{\chi_1^2(\lambda_j)}$ be the cumulative distribution function of $\chi_1^2(\lambda_j)$. It follows that the power of the test under the significance level $\alpha$ and alternative hypothesis $H_{1,j}:\Big|\boldsymbol{L}^T_j\boldsymbol{\beta}\Big|=c$ is
\begin{equation}
	\begin{split}
		\phi= 1 - F_{\chi_{1}^2(\lambda_j)}(\chi_{1, 1-\alpha}^2).
	\end{split}
	\label{powerformula}
\end{equation}

 To ensure sufficient power for each evaluator at a pre-specified alternative hypothesis, we can first fix the power $\phi$ of the tests, and solve Equation (\ref{powerformula}) to obtain the corresponding significance levels $\alpha_j(\phi)$ for rejecting the null hypothesis $H_{0,j}:\boldsymbol{L}^T_j\boldsymbol{\beta}=0$. Under the same power and alternative hypothesis, each evaluator has an evaluator-specific significance level instead of a unified one due to the differences in the estimated variances of the coefficient estimates.

\subsection*{False Discovery Rate Estimation}\label{secfdr}

 \noindent The null hypotheses that we are testing are $H_{0,1}, H_{0,2},\ldots,H_{0,M}$. Due to multiple testing, using a traditional significance level such as 0.05 in each test may lead to a high rate of finding `outlier' evaluators even if they are `normal' ones (i.e. making false discoveries) \cite{benjamini1995controlling, benjamini2001controlling}. In our setting, since the evaluator-specific significance levels are determined by ensuring a pre-specified power of the tests, we are more likely to make false discoveries than the traditional $\alpha$-level hypothesis tests when the pre-specified power is large. To protect us from falsely classifying too many `normal' evaluators as `outliers', we propose to adopt the concept of the false discovery rate (FDR) \cite{benjamini1995controlling} to control the rate of making false positive decisions.

 Let $\boldsymbol{R}$ denote the total number of null hypotheses being rejected (i.e. discoveries) among $H_{0,1},\ldots,H_{0,M}$ and $\boldsymbol{V}$ denote the number of rejected true null hypotheses (i.e. false discoveries). Define the ratio \cite{benjamini1995controlling, benjamini2001control} as
\begin{equation*}
	\boldsymbol{Q}=\begin{cases}
		\frac{\boldsymbol{V}}{\boldsymbol{R}}, &\text{If } \boldsymbol{R}>0,\\
		0, & \text{If } \boldsymbol{R}=0.
	\end{cases}
\end{equation*}
Then, the FDR is the expectation of false discoveries among discoveries: $\text{E}(\boldsymbol{Q})$. Therefore, in the context of our paper, given a power $\phi$, the FDR is:
\begin{equation*}
	\text{E}(\boldsymbol{Q};\phi)=\text{E}\left(\frac{\boldsymbol{V}(\phi)}{\boldsymbol{R}(\phi)}\right).
\end{equation*}
Note that we use the notations $\text{E}(\boldsymbol{Q};\phi)$, $\boldsymbol{V}(\phi)$, $\boldsymbol{R}(\phi)$ to indicate that they are dependent on the pre-specified power $\phi$ of the test. 

 Storey and Tibshirani
\cite{storey2003statistical} provided an approximation formula for the FDR:
\begin{equation}
	\text{E}(\boldsymbol{Q};\phi)=\text{E}\left(\frac{\boldsymbol{V}(\phi)}{\boldsymbol{R}(\phi)}\right)\approx \frac{\text{E}(\boldsymbol{V}(\phi))}{\text{E}(\boldsymbol{R}(\phi))}.
	\label{FDR_approx}
\end{equation}
Recall that given a particular power $\phi$, the corresponding significance levels for hypotheses $H_{0,1}, H_{0,2},\ldots, H_{0,M}$ are $\alpha_1(\phi), \alpha_2(\phi),\ldots, \alpha_M(\phi)$, respectively. Since $\alpha_j(\phi)$ represents the probability of falsely rejecting $H_{0,j}$ given it is true, the numerator $\text{E}(\boldsymbol{V}(\phi))$ in (\ref{FDR_approx}) can be written as
\begin{equation*}
	\text{E}(\boldsymbol{V}(\phi))= \sum_{j\in\mathcal{T}}\alpha_j(\phi),
	\label{nume.exact}
\end{equation*}
where $\mathcal{T}$ is the set of the indexes of the true null hypotheses (i.e. `normal' evaluators). However, the set of true null hypotheses are unknown. We make the assumption of rare `outlier' evaluators; that is $\sum_{j\in\mathcal{T}}\alpha_j(\phi)\approx\sum_{j=1}^M\alpha_j(\phi)$. Therefore, the numerator of (\ref{FDR_approx}) can be approximated by
\begin{equation}
	\text{E}(\boldsymbol{V}(\phi))\approx \sum_{j=1}^{M}\alpha_j(\phi).
	\label{nume}
\end{equation}

 A simple estimate of the denominator $\text{E}(\boldsymbol{R}(\phi))$ is the observed total number of rejected null hypotheses from an experiment \cite{storey2003statistical}; that is,
\begin{equation}
	\text{E}(\boldsymbol{R}(\phi))\approx \boldsymbol{R}(\phi)=\sum_{j=1}^{M}\text{I}(p_j<\alpha_j(\phi)),
	\label{denom}
\end{equation}
where $p_j$ is the $p$-value corresponding to the $j$-th evaluator.

 Plugging (\ref{nume}) and (\ref{denom}) into Formula (\ref{FDR_approx}), we have
\begin{equation}
	\widehat{\text{E}}(\boldsymbol{Q};\phi)=\frac{\sum_{j=1}^{M}\alpha_j(\phi)}{\sum_{j=1}^{M}\text{I}(p_j<\alpha_j(\phi))}.
	\label{final_fdr_est}
\end{equation}

 Note that, in our approach, instead of using a unified significance level for all tests, such as $\alpha=0.05$, each null hypothesis has its own evaluator-specific significance level such that a pre-specified power for detecting a pre-specified alternative hypothesis is achieved for all the hypothesis tests. The estimated FDR, $\widehat{\text{E}}(\boldsymbol{Q};
\phi)$, on the other hand, can inform us of the number of false discoveries that may be made. Therefore, when choosing an appropriate set of significance levels, apart from ensuring sufficient power for the tests, the estimated FDR can be used as another criterion reflecting our tolerance towards making false discoveries.

\subsection*{FDR vs. Power Decision Plot}\label{vsplot}
 \noindent As described in previous sections, for a given power, we could solve Equation (\ref{powerformula}) to get the corresponding evaluator-specific significance levels for rejecting the null hypotheses $H_{0,j}, j=1,\ldots, M$, and based on these significance levels, the corresponding FDR can be estimated using Equation (\ref{final_fdr_est}). Therefore, the relationship between power and FDR can be reflected by a decision plot where the power ($\phi$) is on the x-axis, and the corresponding estimated FDR ($\widehat{\text{E}}(\boldsymbol{Q},\phi)$) is on the y-axis. Based on the decision plot, we can pick up the significance levels at which an acceptable trade-off between power and the FDR is achieved. 

 We could also first select a relatively low FDR and find the corresponding power along with the evaluator-specific significance levels from the decision plot; we can then reject the null hypotheses with $p$-values of the tests less than the thresholds. Alternatively, if we are less concerned about making false discoveries but would like to be able to detect as many potential `outlier' evaluators as possible, then we could first specify a relatively large power, and reject the null hypotheses by comparing the $p$-values with the corresponding evaluator-specific significance levels; the estimated FDR from the decision plot can inform us of the number of false discoveries we might have made.

\subsection*{FDR-based Adjustment}\label{fdr_driven_adjust}

 \noindent We may further adjust the set of rejected null hypotheses based on the estimated FDR, especially when $\widehat{\text{E}}(\boldsymbol{Q};\widetilde{\phi})$ is large under the chosen power $\widetilde{\phi}$.

 Let $\mathcal{R}$ be the set of the rejected null hypotheses, and $k$ be the number of hypotheses in $\mathcal{R}$. Denote the rejected hypotheses as ${H}_{0,(1)}, {H}_{0,(2)}, \ldots, {H}_{0,(k)}$, where they are ordered by their $p$-values in an ascending order. Since $\widehat{\text{E}}(\boldsymbol{Q};\widetilde{\phi})\times k$ approximates the expected number of true null hypotheses that are falsely rejected among ${H}_{0,(1)}, {H}_{0,(2)}, \ldots, {H}_{0,(k)}$, an ad hoc approach to further adjust the rejected null hypotheses based on the estimated FDR is to move the latter $\lceil \widehat{\text{E}}(\boldsymbol{Q};\widetilde{\beta}^p)\times k\rceil$ null hypotheses $H_{0,(k-\lceil \widehat{\text{E}}(\boldsymbol{Q};\widetilde{\beta}^p)\times k\rceil+1)},\ldots, H_{0,(k)}$ out of set $\mathcal{R}$, where $\lceil x\rceil$ rounds $x$ to the nearest integer. Finally we would only reject $H_{0,(1)}, H_{0,(2)},\ldots, H_{0,(k-\lceil \widehat{\text{E}}(\boldsymbol{Q};\widetilde{\beta}^p)\times k\rceil)}$, and the corresponding `outliers' are evaluators $(1), (2),\ldots, \text{ and } (k-\lceil \widehat{\text{E}}(\boldsymbol{Q};\widetilde{\beta}^p)\times k\rceil)$. {\color{black}{Algorithm \ref{alg} summaries the complete quality control procedure.}} 

\RestyleAlgo{ruled}

\begin{algorithm}
	\caption{`Outlier' Detection}\label{alg}
	\textbf{Input}: Measurements from evaluators, with study participants' characteristics data\;
	\textbf{Output}: Potential `Outlier' evaluators \;
	\textbf{First stage}: Run the no-intercept regression: $	\text{E}(Y_i|\boldsymbol{X}_i, \text{T}_i^{(1)},\ldots, \text{T}_i^{(M)} )=\sum_{j=1}^M\beta_j\text{T}_i^{(j)} + \boldsymbol{\gamma}^T\boldsymbol{X}_i$\; 
	
	\textbf{Start of the second stage:} Let $\Phi$ be a series of power values over the range of $(0,1)$, e.g., $\Phi=\texttt{seq(from = 0.1, to = 0.95, by = 0.01)}$
	
\For{each $\phi$ in $\Phi$}{
	\For{j in $1:M$}{
		Derive the significance level $\alpha_j(\phi)$ corresponding to the test for $\widehat{\beta}_j$, based on either the truncated or untruncated test
	}
	Calculate the estimated FDR: $\widehat{\text{E}}(\boldsymbol{Q};\phi)=\frac{\sum_{j=1}^{M}\alpha_j(\phi)}{\sum_{j=1}^{M}\text{I}(p_j<\alpha_j(\phi))}$
}
Create the FDR vs. Power plot by plotting $\phi$ versus the estimated FDR, and determine a reasonable power $\widetilde{\phi}$ based on the plot.

\For{j in $1:M$}{
	\If{p-value $p_j<\alpha_{j}(\tilde{\phi})$}{
		Declare the $j$-th evaluator as an `outlier'}
}

\If{FDR-based adjustment = YES}{
	Let $k=\sum_{j=1}^M\text{I}(p_j<\alpha_j(\widetilde{\phi}))$ and $\widehat{FDR}=\frac{\sum_{j=1}^{M}\alpha_j(\widetilde{\phi})}{\sum_{j=1}^{M}\text{I}(p_j<\alpha_j(\widetilde{\phi}))}$\;
	\If{$k*\widehat{FDR}>1$}{
		Sort the declared `outlier' evaluators using p-values in ascending order
		
		\If{p-values of `outliers' are of order $k-\ceil{k*\widehat{FDR}}$ to $k$}{Reclassify them as `normal' evaluators}
		
	}
}
\textbf{End of the second stage}
	
\end{algorithm}

\section*{Simulation}\label{section3}
\noindent We perform a simulation study to assess the proposed quality control procedure for detecting `outlier' evaluators. As a demonstration, we base our simulations on the audiometrically-assessed hearing threshold measurements at 8 kHz that were obtained in the CHEARS AAA in 2014, where 3,568 participants had assessments in both ears that were measured by 68 different licensed audiologists. We evaluate the performance of the proposed FDR estimator in Equation (\ref{final_fdr_est}), as well as true positives (successfully detecting true `outlier' evaluators) and false positives (falsely classifying `normal' evaluators as `outliers') yielded by our quality control method compared with using a traditional and unified significance level such as $\alpha=0.05$ to reject the null hypotheses.

\subsection*{Data generation}
\noindent We first consider the scenario when evaluators measure a single outcome for each study participant. We generate data based on the model below, mimicking the right ear data obtained from the CHEARS AAA:
\begin{equation*}
	\begin{split}
		Y_i=&\gamma_1\text{age}_i+\gamma_2\text{age}_i^2+\gamma_3\text{I}(\text{very good}_i) \\
		&+\gamma_4\text{I}(\text{a little hearing trouble}_i)+\beta_1\text{Audio}_i^{(1)} +\beta_2\text{Audio}_i^{(2)}\\
		&+\ldots+\beta_{M}\text{Audio}_i^{(M)}+\epsilon_i,
	\end{split}
\end{equation*}
where age is generated from a normal distribution with mean 56.6 years and standard deviation (SD) 4.4; we set the `excellent' self-reported hearing status as the reference group and the prevalences of the other two categories `very good' and `a little hearing trouble' were 0.44 and 0.25, respectively. These values are the same as those in the CHEARS AAA. $\text{Audio}_i^{(j)}, j=1,\ldots, M$, is 1 if the hearing test outcome of the $i$-th study participant is measured by the $j$-th audiologist, and 0 otherwise.

 The coefficients corresponding to age, age$^2$, I(very good), and I(a little hearing trouble) are set to be $\gamma_1=-2.7$, $ \gamma_2=0.03$, $\gamma_3=3.3$ and $ \gamma_4=10.3$, same as the point estimates from the regression analysis on the CHEARS data. The number of audiologists $M$ are set to be 100, and each measures the hearing outcomes on 40 study participants. We set the first 8 audiologists as `outliers', and the remaining ones are `normal' audiologists. The coefficients corresponding to the `normal' audiologists are set to be $\beta_9=\beta_{10}=\ldots=\beta_{100}=67$, while for `outlier' audiologists, we set $\beta_1=\beta_2=\ldots=\beta_5=75$ and $\beta_6=\beta_7=\beta_8=70$. Note that, here, five `outlier' audiologists have very different effects on the hearing test outcomes from `normal' audiologists and three `outlier' audiologists are slightly different from `normal' audiologists. The values 75 and 67 are determined by the averages of the estimated regression coefficients in the regression analysis on the CHEARS data for the audiologists in the upper 10th percentile and those between the lower and upper 10th percentiles, respectively. The residual $\epsilon_i$ is assumed to be normally distributed with mean 0 and standard deviation (SD) $\sigma = 8, 10, 12$, respectively.

\subsection*{Simulation results}

\noindent The simulation is performed for 300 replicates. Shown in Figure \ref{one.ear.fdr} are the FDR vs. Power decision plots under different standard deviation (SD) of the residuals. We set the alternative hypothesis as $H_{1,j}:\Big|\boldsymbol{L}^T_{10\%, j}\boldsymbol{\beta}\Big|=5$. The solid curve is the estimated FDR based on Equation (\ref{final_fdr_est}) averaged over the 300 simulation replicates under powers ($\phi$) ranging from 0.1 to 0.95 with step size of 0.01; a loess curve with the default smoothing span 0.75 is fitted to connect the points. The dashed curve is an empirical version of the true FDR, which for each $\phi$, is the ratio of the number of `normal' audiologists (Audiologists 9 - 100) being falsely detected as `outlier' audiologists to the total number of detected `outlier' audiologists, averaged over the 300 simulation replicates. The horizontal dot-dash line is the empirical version of the true FDR if we use $\alpha=0.05$ as the significance level for rejecting the null hypotheses averaged over the 300 simulation replicates. 

 As shown in the decision plot, the estimated FDR is very close to the true FDR when $\sigma=8 \text{ and } 10$; while it slightly overestimate the true value when $\sigma=12$. Moreover, as the SD of the residual increases, the FDR also increases. For example, when $\sigma=8$, the FDR is less than 0.165 under power 0.95, while if $\sigma$ increases to 12, the FDR is greater than 0.8 under the same power. Define the noise ratio as $\frac{\sigma^2}{\text{Var}(Y)}$, which is the proportion of the variance of the residual among the total variance of the outcome measurement. The corresponding noise ratios are approximately 0.52, 0.64, and 0.72 for $\sigma=8, 10 \text{ and }12$. When the noise ratio increases, we are more likely to make false discoveries. Therefore, when performing quality control, including all the possible predictors and confounders in the first stage regression is crucial; this way, we can minimize the residual of the first stage regression and, as a result, minimize the FDR.

 Compared with an approach that uses a fixed significance level $\alpha=0.05$, our method enjoys more flexibility since we can choose the evaluator-specific significance levels by considering both the power and FDR. When $\sigma=8$, under any power, our approach has a much lower FDR than using $\alpha=0.05$ as the threshold; and when $\sigma=10 \text{ and }12$, even though the FDR increases, it is still smaller than the FDR if using $\alpha=0.05$ as the threshold, when the power is chosen to be less than 0.8 and 0.75, respectively. 

 Define the true positive proportion for each true `outlier' audiologist (i.e., Audiologists 1 to 8) as the proportion of simulation replicates that correctly detect the audiologist as an `outlier' over the 300 simulation replicates, and the false positive proportion for each true `normal' audiologist (i.e., Audiologists 9 to 100) as the proportion of simulation replicates that falsely identify the audiologist as an `outlier' over the 300 simulation replicates. Figure \ref{fig2.a} and Figure \ref{fig2.b} show the true positive proportions for Audiologists 1 to 8, and false positive proportions for the `normal' audiologists (For illustration, we select Audiologists 9 to 16.), where $\sigma=8$ when generating the data, and the alternative hypothesis is set as $H_{1,j}:\Big|\boldsymbol{L}_{10\%, j}^T\boldsymbol{\beta}\Big|=5$. The black points are the proportions based on our quality control procedure under different powers of the tests; while the horizontal dotted lines are the proportions calculated using $\alpha=0.05$ as the threshold for rejecting the null hypotheses. We consider both the unadjusted procedure and the FDR-based adjusted procedure. 

 For the unadjusted procedure, as the power increases, the true positive proportions for Audiologists 1 to 5 reach to 1 quickly, which is expected since the difference between their coefficients and those of the `normal' audiologists are set to be 8, greater than the difference used in the alternative hypothesis $H_{1,j}: \Big|\boldsymbol{L}_{10\%, j}^T\boldsymbol{\beta}\Big|=5$. However, for Audiologists 6 to 8, since their coefficients are only 3 larger than the `normal' audiologists, the true positive proportions are far less than 1 even when the power is large. Compared to the approach that uses $\alpha=0.05$ as the threshold, our quality control procedure has smaller true positive proportions when the power of test is smaller than 0.3, 0.6, 0.7 for $\sigma=8, 10, 12$, but gradually they will increase to approximately the same or even higher level. For the `normal' audiologists (Audiologists 9 to 16), the false positive proportions are approximately 0.05 if using $\alpha=0.05$ as the threshold. Our quality control procedure has even smaller false positive proportions when $\sigma=8 \text{ and } 10$ under nearly every power considered. When $\sigma=12$, the false positive proportions are still smaller than those from using $\alpha=0.05$ as the threshold, if the power is no larger than 0.9.

 Compared with the unadjusted procedure, the FDR-based adjusted true positive proportions for the true `outlier' audiologists and false positive proportions for `normal' audiologists do not change much in the case of $\sigma=8$ since the FDR is small, and the adjustment is minor.  As $\sigma$ increases, for example, when $\sigma=10$, the FDR is large enough to yield sufficient number of adjustments for power larger than 0.75. Apart from a decrease in the false positive proportions for the true `normal' audiologists (Audiologists 9 to 16), we also observe a decrease in the true positive proportions for the true `outlier' audiologists (Audiologists 1 to 8). Therefore, the ad hoc FDR-based adjustment helps to reduce the chances of making false discoveries, with a price of a reduction in the probability of making true positive decisions.

Moreover, we also conducted a simulation study for the scenarios when outcomes are correlated. The data generation process and simulation results are presented in Supplementary Material Section 1. The simulation results are similar with the single measurement scenarios; our outlier detection procedure typically has lower false positive proportions for the true `normal' audiologists and higher true positive proportions for the true `outlier' audiologists compared with the approach that fix the significance level at $\alpha=0.05$.

\section*{Application}\label{section4}
\noindent To illustrate our method, we apply our method to detect `outlier' audiologists for the audiometrically-assessed hearing threshold measurements in the CHEARS AAA collected in 2014, when the baseline testing was completed on 3,749 participants. We focus on the test results at 8 kHz. We use the GEE approach in the first stage regression analysis and we include $\text{age}, \text{age}^2$, self-reported hearing status (`excellent', ` very good' and `a little hearing trouble'), and dummy variables for the 68 audiologists in the regression model. This regression is fitted using SAS proc genmod, assuming an exchangeable working variance-covariance structure. 

 We display the scatter plots of $ \widehat{\beta}_i-\frac{1}{M}\sum_{q=1}^{M}\widehat{\beta}_{q}$ and $ \widehat{\beta}_i-\frac{1}{M-2[M\cdot\delta]}\sum_{q=[M\cdot\delta]+1}^{M-[M\cdot\delta]}\widehat{\beta}_{(q)}$, with $M=68, \delta=0.1$, in Figure \ref{gee_ear_plot_mean}. Regardless of whether we are comparing with the untruncated mean or the 10\% truncated mean, the plots are similar. As shown in Figure \ref{fig3.a} and Figure \ref{fig3.b}, Audiologist 13 has a much larger ($>10 \text{ dB}$) coefficient estimate than their counterparts, and Audiologist 4 has a much smaller ($<10 \text{ dB}$) coefficient estimate than the rest of the audiologists. Moreover, Audiologists 14, 15, 22, 47, 48, 54, 55 and 59 have a mildly different (5-10$\text{ dB}$) coefficient estimates from the average effect. 

 Figure \ref{fig4.a} to Figure \ref{fig4.d} show the FDR vs. Power decision plots, where the hypothesis tests are performed to compare each audiologist's regression coefficient with both the untruncated mean and the 10\% truncated mean. We fix the alternative hypothesis as $H_{1,j}:\Big| \boldsymbol{L}^T_{j}\boldsymbol{\beta}\Big|=5 \text{ and } 10$, and $H_{1,j}: \Big|\boldsymbol{L}^T_{10\%, j}\boldsymbol{\beta}\Big|=5 \text{ and } 10$, respectively, for $j=1,\ldots, 68$. Based on the decision plots, `outlier' audiologists can be detected by choosing an appropriate set of significance levels that correspond to reasonable power and FDR. Tables \ref{real.gee.1} and \ref{real.gee.2} summarize the results when setting the power at 0.8 or the estimated FDR at 0.5. As shown in these tables, Audiologists 4 and 13 are detected as `outliers' by all of the approaches regardless of the power, FDR or the alternative hypothesis considered, and Audiologist 48 is detected by all of the approaches under the alternative hypothesis $H_{1,j}: \Big|\boldsymbol{L}_{10\%,j}^T\boldsymbol{\beta}\Big|=5$ and $H_{1,j}: \Big|\boldsymbol{L}_{j}^T\boldsymbol{\beta}\Big|=5$. Therefore, Audiologists 4, 13 and 48 are likely to be `outlier' audiologists, suggesting close scrutiny may be merited. However, for the approach of using $\alpha=0.05$ to reject the null hypotheses as shown in the last two rows of the tables, apart from being not flexible as compared with our method, it also suffers from the problem that the power of tests for different audiologists varies significantly with a minimum of 0.55 and a maximum of 1.00.

\section*{Discussion}\label{section5}

\noindent In this paper, we propose a novel method to address a common issue in large epidemiologic studies that rely on multiple evaluators to obtain exposure or outcome measurements to optimize data quality during data collection stage. Specifically, we developed a two-stage {\color{black}{algorithm}} to detect `outlier' evaluators, who may tend to have higher or lower measurements than those of their counterparts. In the first stage, we fit a regression model for the measurements against evaluators and study participants' characteristics that could predict the measurements. In the second stage, based on the regression coefficients in the first stage, we perform hypothesis tests to compare the mean measurement of each evaluator with the average mean measurements over all evaluators adjusting for the characteristics of the individuals evaluated. Different from the traditional hypothesis testing procedure where controlling type-I error is the primary focus, we also attach equal importance to ensuring an appropriate level of type-II error since our goal is to detect as many potential `outlier' evaluators as possible for quality control purpose. We derive the evaluator-specific significance levels for rejecting the null hypotheses under selected powers of the tests. {\color{black}{These significance levels are not necessarily 0.05 and are different across audiologists due to the differences in the variances of the coefficient estimates.}} To account for the issue of multiple comparisons, we also derive an FDR-estimator. An FDR vs. Power decision plot can be created, and based on this plot, the evaluator-specific significance levels for rejecting the null hypotheses can be determined such that both FDR and Power are acceptable. 

{\color{black}{When performing hypothesis tests to detect `outlier' evaluators, we proposed to compare the coefficient estimates to the truncated mean to prevent those `outlier' evaluators from contaminating the estimated normal effect. Alternatively, we can consider an interval null, that is $H_0: |\beta_i - \frac{1}{M}\sum_{j=1}^{M} \beta_j| \le a$  for some constants $a>0$. A challenge of this method might be how to select $a$. We will consider this method in our future research and compare it with the current method. Moreover, when calculating the evaluator-specific significance level, the knowledge of the alternative hypothesis is needed. However, if the prior knowledge is not available, we recommend performing sensitivity analysis for a series of reasonable values of the alternative hypothesis. In addition, the FDR approximation in Equation (2) holds when the number of hypotheses ($M$) being conducted is large. However, when $M$ is small, alternatively, we can use the Benjamini-Hochberg (BH) procedure to control the FDR \cite{benjamini1995controlling}. The BH procedure proceeds by first specifying an FDR level $\alpha$, and sort the null hypothesis based on p-values in ascending order ($P_{(1)}, P_{(2)},\ldots, P_{(M)}$). Then the largest $k$ such that $P_{(k)}\le \frac{k}{M}\alpha$ is obtained, and the first $k$ null hypotheses will be rejected. The BH procedure can ensure that the FDR is controlled at level $\alpha$. However, different from our approach, the BH procedure does not consider the power of tests and to be conservative, we might use a relatively larger $\alpha$ level such as 0.1 when conducting the BH procedure.}}

 There are several important points for consideration based on our work. First, an increase in the noise ratio $\frac{\sigma^2}{\text{Var}(Y)}$ will increase FDR, especially when the power of the test is large. Therefore, in the first stage regression, it is crucial to include all potential predictors of the measurements as regressors. Second, the proposed method assumes that the evaluator effect on the measurements is not modified by the participants' characteristics. In the case when this assumption is violated, we can estimate the evaluator effect in each category of the potential effect modifier by including the evaluator indicator-effect modifier interactions in the first stage regression model, and then we can regard the same evaluator for testing study participants in different categories of the effect modifier as if they were different evaluators. This way, an evaluator could be detected as an `outlier' only when testing study participants in a specific category of the effect modifier. Third, to accommodate situations where the measurements are not continuous, a link function can be used in the first stage regression, such as the logit link for binary measurements, and log link for count measurements.

 The regular regression and GEE approach may not lead to reliable $\beta$-estimator if the numbers of study participants tested by some evaluators are small. In this case, an alternative method is to treat the measurements from the same evaluator as a cluster and to use the mixed effects model in the first stage regression analysis. In the scenario where each participant has a single measurement, this mixed effects model may include an evaluator-specific random intercept in addition to the fixed effect participants' characteristics; the estimated value of the $j$-th evaluator-specific intercept is $\hat{\beta}_j$. Similarly, in the scenario where the participants have multiple measurements, the mixed effects model may include both evaluators and participants (nested within evaluator) as random effects. Once the mixed effects model obtains $\widehat{\boldsymbol{\beta}}$ and $\widehat{Var}(\widehat{\boldsymbol{\beta}})$, the rest of the methods are the same as those stated in Subsection `Hypothesis Testing' to Subsection `FDR-based Adjustment'  of this paper.

 In additional to being useful during the data collection stage of epidemiologic studies, our outlier detection method can also be valuable in the clinical setting for the detection of `outlier' evaluators (e.g. health providers or technicians); since treatment prescriptions often rely on measurements from evaluators, and inaccurate measurements may lead to incorrect treatment decisions. Further, our method can be used to perform sensitivity analyses for statistical analysis procedures. For example, for studies based on laboratory measurements of biomarkers such as plasma or urine metabolites that are measured in different batches, our method can help to identify potential `outlier' batches, and a sensitivity analysis can be conducted by excluding those `outlier' batches to re-estimate the parameters of interests. 

 R code for implementing the proposed method is available at \url{https://www.hsph.harvard.edu/molin-wang/software}.

{\color{black}{\section*{Conclusions}
		Our two-stage algorithm is a useful method for detecting `outlier' evaluators who tend to give higher or lower measurements than their counterparts after adjusting for study participants' characteristics. Compared with traditional hypothesis tests that focus on type-I error, we also attach importance to the type-II error so that as many potential `outliers' can be identified, and an estimated FDR is used to control for the false positive rate. We recommend applying our method for `outlier' detection during data collection stage to improve data quality.}}


\begin{backmatter}
\section*{Declarations}

		\section*{Ethics approval and consent to participate}
Not applicable.

		\section*{Consent for publication}
Not applicable.

		\section*{Availability of data and materials}
The data that support the findings of this study are available from Nurses' Health Study (NHS) II but restrictions apply to the availability of these data, which were used under license for the current study, and so are not publicly available. Data are however available from the authors upon reasonable request and with permission of Nurses' Health Study (NHS) II.

		\section*{Competing interests}
The authors declare that they have no competing interests.

		\section*{Funding}
This work is supported by NIH grant R01DC017717.

		\section*{Authors' contributions}
YW, BR and MW developed the methods; YW designed and conducted the simulation study, wrote the first draft of the manuscript. SC, BR, GC,  and MW reviewed the manuscript critically. All authors read and approved the final manuscript.

\section*{Acknowledgements}
{\color{black}{We are thankful to the study participants in CHEARS.
	
			\section*{Abbreviations}
	FDR: False discovery rate; CHEARS: Conservation of Hearing Study; AAA: Audiology Assessment Arm; NHS: Nurses' Health Study; GEE: Generalized Estimating Equations

}}


\bibliographystyle{bmc-mathphys} 
\bibliography{Reference}      


\begin{thebibliography}{18}
\ifx \bisbn   \undefined \def \bisbn  #1{ISBN #1}\fi
\ifx \binits  \undefined \def \binits#1{#1}\fi
\ifx \bauthor  \undefined \def \bauthor#1{#1}\fi
\ifx \batitle  \undefined \def \batitle#1{#1}\fi
\ifx \bjtitle  \undefined \def \bjtitle#1{#1}\fi
\ifx \bvolume  \undefined \def \bvolume#1{\textbf{#1}}\fi
\ifx \byear  \undefined \def \byear#1{#1}\fi
\ifx \bissue  \undefined \def \bissue#1{#1}\fi
\ifx \bfpage  \undefined \def \bfpage#1{#1}\fi
\ifx \blpage  \undefined \def \blpage #1{#1}\fi
\ifx \burl  \undefined \def \burl#1{\textsf{#1}}\fi
\ifx \doiurl  \undefined \def \doiurl#1{\textsf{#1}}\fi
\ifx \betal  \undefined \def \betal{\textit{et al.}}\fi
\ifx \binstitute  \undefined \def \binstitute#1{#1}\fi
\ifx \binstitutionaled  \undefined \def \binstitutionaled#1{#1}\fi
\ifx \bctitle  \undefined \def \bctitle#1{#1}\fi
\ifx \beditor  \undefined \def \beditor#1{#1}\fi
\ifx \bpublisher  \undefined \def \bpublisher#1{#1}\fi
\ifx \bbtitle  \undefined \def \bbtitle#1{#1}\fi
\ifx \bedition  \undefined \def \bedition#1{#1}\fi
\ifx \bseriesno  \undefined \def \bseriesno#1{#1}\fi
\ifx \blocation  \undefined \def \blocation#1{#1}\fi
\ifx \bsertitle  \undefined \def \bsertitle#1{#1}\fi
\ifx \bsnm \undefined \def \bsnm#1{#1}\fi
\ifx \bsuffix \undefined \def \bsuffix#1{#1}\fi
\ifx \bparticle \undefined \def \bparticle#1{#1}\fi
\ifx \barticle \undefined \def \barticle#1{#1}\fi
\ifx \bconfdate \undefined \def \bconfdate #1{#1}\fi
\ifx \botherref \undefined \def \botherref #1{#1}\fi
\ifx \url \undefined \def \url#1{\textsf{#1}}\fi
\ifx \bchapter \undefined \def \bchapter#1{#1}\fi
\ifx \bbook \undefined \def \bbook#1{#1}\fi
\ifx \bcomment \undefined \def \bcomment#1{#1}\fi
\ifx \oauthor \undefined \def \oauthor#1{#1}\fi
\ifx \citeauthoryear \undefined \def \citeauthoryear#1{#1}\fi
\ifx \endbibitem  \undefined \def \endbibitem {}\fi
\ifx \bconflocation  \undefined \def \bconflocation#1{#1}\fi
\ifx \arxivurl  \undefined \def \arxivurl#1{\textsf{#1}}\fi
\csname PreBibitemsHook\endcsname

\bibitem{cruickshanks1998prevalence}
\begin{barticle}
\bauthor{\bsnm{Cruickshanks}, \binits{K.J.}},
\bauthor{\bsnm{Wiley}, \binits{T.L.}},
\bauthor{\bsnm{Tweed}, \binits{T.S.}},
\bauthor{\bsnm{Klein}, \binits{B.E.}},
\bauthor{\bsnm{Klein}, \binits{R.}},
\bauthor{\bsnm{Mares-Perlman}, \binits{J.A.}},
\bauthor{\bsnm{Nondahl}, \binits{D.M.}}:
\batitle{Prevalence of hearing loss in older adults in beaver dam, wisconsin:
  The epidemiology of hearing loss study}.
\bjtitle{American journal of epidemiology}
\bvolume{148}(\bissue{9}),
\bfpage{879}--\blpage{886}
(\byear{1998})
\end{barticle}
\endbibitem

\bibitem{shargorodsky2010change}
\begin{barticle}
\bauthor{\bsnm{Shargorodsky}, \binits{J.}},
\bauthor{\bsnm{Curhan}, \binits{S.G.}},
\bauthor{\bsnm{Curhan}, \binits{G.C.}},
\bauthor{\bsnm{Eavey}, \binits{R.}}:
\batitle{Change in prevalence of hearing loss in us adolescents}.
\bjtitle{Jama}
\bvolume{304}(\bissue{7}),
\bfpage{772}--\blpage{778}
(\byear{2010})
\end{barticle}
\endbibitem

\bibitem{gopinath2010incidence}
\begin{barticle}
\bauthor{\bsnm{Gopinath}, \binits{B.}},
\bauthor{\bsnm{McMahon}, \binits{C.M.}},
\bauthor{\bsnm{Rochtchina}, \binits{E.}},
\bauthor{\bsnm{Karpa}, \binits{M.J.}},
\bauthor{\bsnm{Mitchell}, \binits{P.}}:
\batitle{Incidence, persistence, and progression of tinnitus symptoms in older
  adults: the blue mountains hearing study}.
\bjtitle{Ear and hearing}
\bvolume{31}(\bissue{3}),
\bfpage{407}--\blpage{412}
(\byear{2010})
\end{barticle}
\endbibitem

\bibitem{zhang2013association}
\begin{barticle}
\bauthor{\bsnm{Zhang}, \binits{X.}},
\bauthor{\bsnm{Bullard}, \binits{K.M.}},
\bauthor{\bsnm{Cotch}, \binits{M.F.}},
\bauthor{\bsnm{Wilson}, \binits{M.R.}},
\bauthor{\bsnm{Rovner}, \binits{B.W.}},
\bauthor{\bsnm{McGwin}, \binits{G.}},
\bauthor{\bsnm{Owsley}, \binits{C.}},
\bauthor{\bsnm{Barker}, \binits{L.}},
\bauthor{\bsnm{Crews}, \binits{J.E.}},
\bauthor{\bsnm{Saaddine}, \binits{J.B.}}:
\batitle{Association between depression and functional vision loss in persons
  20 years of age or older in the united states, nhanes 2005-2008}.
\bjtitle{JAMA ophthalmology}
\bvolume{131}(\bissue{5}),
\bfpage{573}--\blpage{581}
(\byear{2013})
\end{barticle}
\endbibitem

\bibitem{klein2014relation}
\begin{barticle}
\bauthor{\bsnm{Klein}, \binits{R.}},
\bauthor{\bsnm{Lee}, \binits{K.E.}},
\bauthor{\bsnm{Gangnon}, \binits{R.E.}},
\bauthor{\bsnm{Klein}, \binits{B.E.}}:
\batitle{Relation of smoking, drinking, and physical activity to changes in
  vision over a 20-year period: the beaver dam eye study}.
\bjtitle{Ophthalmology}
\bvolume{121}(\bissue{6}),
\bfpage{1220}--\blpage{1228}
(\byear{2014})
\end{barticle}
\endbibitem

\bibitem{mccullough2019circulating}
\begin{barticle}
\bauthor{\bsnm{McCullough}, \binits{M.L.}},
\bauthor{\bsnm{Zoltick}, \binits{E.S.}},
\bauthor{\bsnm{Weinstein}, \binits{S.J.}},
\bauthor{\bsnm{Fedirko}, \binits{V.}},
\bauthor{\bsnm{Wang}, \binits{M.}},
\bauthor{\bsnm{Cook}, \binits{N.R.}},
\bauthor{\bsnm{Eliassen}, \binits{A.H.}},
\bauthor{\bsnm{Zeleniuch-Jacquotte}, \binits{A.}},
\bauthor{\bsnm{Agnoli}, \binits{C.}},
\bauthor{\bsnm{Albanes}, \binits{D.}}, \betal:
\batitle{Circulating vitamin d and colorectal cancer risk: an international
  pooling project of 17 cohorts}.
\bjtitle{JNCI: Journal of the National Cancer Institute}
\bvolume{111}(\bissue{2}),
\bfpage{158}--\blpage{169}
(\byear{2019})
\end{barticle}
\endbibitem

\bibitem{carroll2006measurement}
\begin{bbook}
\bauthor{\bsnm{Carroll}, \binits{R.J.}},
\bauthor{\bsnm{Ruppert}, \binits{D.}},
\bauthor{\bsnm{Stefanski}, \binits{L.A.}},
\bauthor{\bsnm{Crainiceanu}, \binits{C.M.}}:
\bbtitle{Measurement Error in Nonlinear Models: a Modern Perspective}.
\bpublisher{Chapman and Hall/CRC}, \blocation{???}
(\byear{2006})
\end{bbook}
\endbibitem

\bibitem{curhan2018adherence}
\begin{barticle}
\bauthor{\bsnm{Curhan}, \binits{S.G.}},
\bauthor{\bsnm{Wang}, \binits{M.}},
\bauthor{\bsnm{Eavey}, \binits{R.D.}},
\bauthor{\bsnm{Stampfer}, \binits{M.J.}},
\bauthor{\bsnm{Curhan}, \binits{G.C.}}:
\batitle{Adherence to healthful dietary patterns is associated with lower risk
  of hearing loss in women}.
\bjtitle{The Journal of nutrition}
\bvolume{148}(\bissue{6}),
\bfpage{944}--\blpage{951}
(\byear{2018})
\end{barticle}
\endbibitem

\bibitem{curhan2020prospective}
\begin{barticle}
\bauthor{\bsnm{Curhan}, \binits{S.G.}},
\bauthor{\bsnm{Halpin}, \binits{C.}},
\bauthor{\bsnm{Wang}, \binits{M.}},
\bauthor{\bsnm{Eavey}, \binits{R.D.}},
\bauthor{\bsnm{Curhan}, \binits{G.C.}}:
\batitle{Prospective study of dietary patterns and hearing threshold
  elevation}.
\bjtitle{American Journal of Epidemiology}
\bvolume{189}(\bissue{3}),
\bfpage{204}--\blpage{214}
(\byear{2020})
\end{barticle}
\endbibitem

\bibitem{liang1986longitudinal}
\begin{barticle}
\bauthor{\bsnm{Liang}, \binits{K.-Y.}},
\bauthor{\bsnm{Zeger}, \binits{S.L.}}:
\batitle{Longitudinal data analysis using generalized linear models}.
\bjtitle{Biometrika}
\bvolume{73}(\bissue{1}),
\bfpage{13}--\blpage{22}
(\byear{1986})
\end{barticle}
\endbibitem

\bibitem{zeger1986longitudinal}
\begin{botherref}
\oauthor{\bsnm{Zeger}, \binits{S.L.}},
\oauthor{\bsnm{Liang}, \binits{K.-Y.}}:
Longitudinal data analysis for discrete and continuous outcomes.
Biometrics,
121--130
(1986)
\end{botherref}
\endbibitem

\bibitem{harrell2015regression}
\begin{bbook}
\bauthor{\bsnm{Harrell~Jr}, \binits{F.E.}}:
\bbtitle{Regression Modeling Strategies: with Applications to Linear Models,
  Logistic and Ordinal Regression, and Survival Analysis}.
\bpublisher{Springer}, \blocation{???}
(\byear{2015})
\end{bbook}
\endbibitem

\bibitem{wilcox2011introduction}
\begin{bbook}
\bauthor{\bsnm{Wilcox}, \binits{R.R.}}:
\bbtitle{Introduction to Robust Estimation and Hypothesis Testing}.
\bpublisher{Academic press}, \blocation{???}
(\byear{2011})
\end{bbook}
\endbibitem

\bibitem{lehmann2006testing}
\begin{bbook}
\bauthor{\bsnm{Lehmann}, \binits{E.L.}},
\bauthor{\bsnm{Romano}, \binits{J.P.}}:
\bbtitle{Testing Statistical Hypotheses}.
\bpublisher{Springer}, \blocation{???}
(\byear{2006})
\end{bbook}
\endbibitem

\bibitem{benjamini1995controlling}
\begin{barticle}
\bauthor{\bsnm{Benjamini}, \binits{Y.}},
\bauthor{\bsnm{Hochberg}, \binits{Y.}}:
\batitle{Controlling the false discovery rate: a practical and powerful
  approach to multiple testing}.
\bjtitle{Journal of the Royal statistical society: series B (Methodological)}
\bvolume{57}(\bissue{1}),
\bfpage{289}--\blpage{300}
(\byear{1995})
\end{barticle}
\endbibitem

\bibitem{benjamini2001controlling}
\begin{barticle}
\bauthor{\bsnm{Benjamini}, \binits{Y.}},
\bauthor{\bsnm{Drai}, \binits{D.}},
\bauthor{\bsnm{Elmer}, \binits{G.}},
\bauthor{\bsnm{Kafkafi}, \binits{N.}},
\bauthor{\bsnm{Golani}, \binits{I.}}:
\batitle{Controlling the false discovery rate in behavior genetics research}.
\bjtitle{Behavioural brain research}
\bvolume{125}(\bissue{1-2}),
\bfpage{279}--\blpage{284}
(\byear{2001})
\end{barticle}
\endbibitem

\bibitem{benjamini2001control}
\begin{botherref}
\oauthor{\bsnm{Benjamini}, \binits{Y.}},
\oauthor{\bsnm{Yekutieli}, \binits{D.}}:
The control of the false discovery rate in multiple testing under dependency.
Annals of statistics,
1165--1188
(2001)
\end{botherref}
\endbibitem

\bibitem{storey2003statistical}
\begin{barticle}
\bauthor{\bsnm{Storey}, \binits{J.D.}},
\bauthor{\bsnm{Tibshirani}, \binits{R.}}:
\batitle{Statistical significance for genomewide studies}.
\bjtitle{Proceedings of the National Academy of Sciences}
\bvolume{100}(\bissue{16}),
\bfpage{9440}--\blpage{9445}
(\byear{2003})
\end{barticle}
\endbibitem

\end{thebibliography}

\newcommand{\BMCxmlcomment}[1]{}

\BMCxmlcomment{

<refgrp>

<bibl id="B1">
  <title><p>Prevalence of hearing loss in older adults in Beaver Dam,
  Wisconsin: The epidemiology of hearing loss study</p></title>
  <aug>
    <au><snm>Cruickshanks</snm><fnm>KJ</fnm></au>
    <au><snm>Wiley</snm><fnm>TL</fnm></au>
    <au><snm>Tweed</snm><fnm>TS</fnm></au>
    <au><snm>Klein</snm><fnm>BE</fnm></au>
    <au><snm>Klein</snm><fnm>R</fnm></au>
    <au><snm>Mares Perlman</snm><fnm>JA</fnm></au>
    <au><snm>Nondahl</snm><fnm>DM</fnm></au>
  </aug>
  <source>American journal of epidemiology</source>
  <publisher>Oxford University Press</publisher>
  <pubdate>1998</pubdate>
  <volume>148</volume>
  <issue>9</issue>
  <fpage>879</fpage>
  <lpage>-886</lpage>
</bibl>

<bibl id="B2">
  <title><p>Change in prevalence of hearing loss in US adolescents</p></title>
  <aug>
    <au><snm>Shargorodsky</snm><fnm>J</fnm></au>
    <au><snm>Curhan</snm><fnm>SG</fnm></au>
    <au><snm>Curhan</snm><fnm>GC</fnm></au>
    <au><snm>Eavey</snm><fnm>R</fnm></au>
  </aug>
  <source>Jama</source>
  <publisher>American Medical Association</publisher>
  <pubdate>2010</pubdate>
  <volume>304</volume>
  <issue>7</issue>
  <fpage>772</fpage>
  <lpage>-778</lpage>
</bibl>

<bibl id="B3">
  <title><p>Incidence, persistence, and progression of tinnitus symptoms in
  older adults: the Blue Mountains Hearing Study</p></title>
  <aug>
    <au><snm>Gopinath</snm><fnm>B</fnm></au>
    <au><snm>McMahon</snm><fnm>CM</fnm></au>
    <au><snm>Rochtchina</snm><fnm>E</fnm></au>
    <au><snm>Karpa</snm><fnm>MJ</fnm></au>
    <au><snm>Mitchell</snm><fnm>P</fnm></au>
  </aug>
  <source>Ear and hearing</source>
  <publisher>LWW</publisher>
  <pubdate>2010</pubdate>
  <volume>31</volume>
  <issue>3</issue>
  <fpage>407</fpage>
  <lpage>-412</lpage>
</bibl>

<bibl id="B4">
  <title><p>Association between depression and functional vision loss in
  persons 20 years of age or older in the United States, NHANES
  2005-2008</p></title>
  <aug>
    <au><snm>Zhang</snm><fnm>X</fnm></au>
    <au><snm>Bullard</snm><fnm>KM</fnm></au>
    <au><snm>Cotch</snm><fnm>MF</fnm></au>
    <au><snm>Wilson</snm><fnm>MR</fnm></au>
    <au><snm>Rovner</snm><fnm>BW</fnm></au>
    <au><snm>McGwin</snm><fnm>G</fnm></au>
    <au><snm>Owsley</snm><fnm>C</fnm></au>
    <au><snm>Barker</snm><fnm>L</fnm></au>
    <au><snm>Crews</snm><fnm>JE</fnm></au>
    <au><snm>Saaddine</snm><fnm>JB</fnm></au>
  </aug>
  <source>JAMA ophthalmology</source>
  <publisher>American Medical Association</publisher>
  <pubdate>2013</pubdate>
  <volume>131</volume>
  <issue>5</issue>
  <fpage>573</fpage>
  <lpage>-581</lpage>
</bibl>

<bibl id="B5">
  <title><p>Relation of smoking, drinking, and physical activity to changes in
  vision over a 20-year period: the Beaver Dam Eye Study</p></title>
  <aug>
    <au><snm>Klein</snm><fnm>R</fnm></au>
    <au><snm>Lee</snm><fnm>KE</fnm></au>
    <au><snm>Gangnon</snm><fnm>RE</fnm></au>
    <au><snm>Klein</snm><fnm>BE</fnm></au>
  </aug>
  <source>Ophthalmology</source>
  <publisher>Elsevier</publisher>
  <pubdate>2014</pubdate>
  <volume>121</volume>
  <issue>6</issue>
  <fpage>1220</fpage>
  <lpage>-1228</lpage>
</bibl>

<bibl id="B6">
  <title><p>Circulating vitamin D and colorectal cancer risk: an international
  pooling project of 17 cohorts</p></title>
  <aug>
    <au><snm>McCullough</snm><fnm>ML</fnm></au>
    <au><snm>Zoltick</snm><fnm>ES</fnm></au>
    <au><snm>Weinstein</snm><fnm>SJ</fnm></au>
    <au><snm>Fedirko</snm><fnm>V</fnm></au>
    <au><snm>Wang</snm><fnm>M</fnm></au>
    <au><snm>Cook</snm><fnm>NR</fnm></au>
    <au><snm>Eliassen</snm><fnm>AH</fnm></au>
    <au><snm>Zeleniuch Jacquotte</snm><fnm>A</fnm></au>
    <au><snm>Agnoli</snm><fnm>C</fnm></au>
    <au><snm>Albanes</snm><fnm>D</fnm></au>
    <au><cnm>others</cnm></au>
  </aug>
  <source>JNCI: Journal of the National Cancer Institute</source>
  <publisher>Narnia</publisher>
  <pubdate>2019</pubdate>
  <volume>111</volume>
  <issue>2</issue>
  <fpage>158</fpage>
  <lpage>-169</lpage>
</bibl>

<bibl id="B7">
  <title><p>Measurement error in nonlinear models: a modern
  perspective</p></title>
  <aug>
    <au><snm>Carroll</snm><fnm>RJ</fnm></au>
    <au><snm>Ruppert</snm><fnm>D</fnm></au>
    <au><snm>Stefanski</snm><fnm>LA</fnm></au>
    <au><snm>Crainiceanu</snm><fnm>CM</fnm></au>
  </aug>
  <publisher>Chapman and Hall/CRC</publisher>
  <pubdate>2006</pubdate>
</bibl>

<bibl id="B8">
  <title><p>Adherence to healthful dietary patterns is associated with lower
  risk of hearing loss in women</p></title>
  <aug>
    <au><snm>Curhan</snm><fnm>SG</fnm></au>
    <au><snm>Wang</snm><fnm>M</fnm></au>
    <au><snm>Eavey</snm><fnm>RD</fnm></au>
    <au><snm>Stampfer</snm><fnm>MJ</fnm></au>
    <au><snm>Curhan</snm><fnm>GC</fnm></au>
  </aug>
  <source>The Journal of nutrition</source>
  <publisher>Oxford University Press</publisher>
  <pubdate>2018</pubdate>
  <volume>148</volume>
  <issue>6</issue>
  <fpage>944</fpage>
  <lpage>-951</lpage>
</bibl>

<bibl id="B9">
  <title><p>Prospective Study of Dietary Patterns and Hearing Threshold
  Elevation</p></title>
  <aug>
    <au><snm>Curhan</snm><fnm>SG</fnm></au>
    <au><snm>Halpin</snm><fnm>C</fnm></au>
    <au><snm>Wang</snm><fnm>M</fnm></au>
    <au><snm>Eavey</snm><fnm>RD</fnm></au>
    <au><snm>Curhan</snm><fnm>GC</fnm></au>
  </aug>
  <source>American Journal of Epidemiology</source>
  <publisher>Oxford University Press</publisher>
  <pubdate>2020</pubdate>
  <volume>189</volume>
  <issue>3</issue>
  <fpage>204</fpage>
  <lpage>-214</lpage>
</bibl>

<bibl id="B10">
  <title><p>Longitudinal data analysis using generalized linear
  models</p></title>
  <aug>
    <au><snm>Liang</snm><fnm>KY</fnm></au>
    <au><snm>Zeger</snm><fnm>SL</fnm></au>
  </aug>
  <source>Biometrika</source>
  <publisher>Oxford University Press</publisher>
  <pubdate>1986</pubdate>
  <volume>73</volume>
  <issue>1</issue>
  <fpage>13</fpage>
  <lpage>-22</lpage>
</bibl>

<bibl id="B11">
  <title><p>Longitudinal data analysis for discrete and continuous
  outcomes</p></title>
  <aug>
    <au><snm>Zeger</snm><fnm>SL</fnm></au>
    <au><snm>Liang</snm><fnm>KY</fnm></au>
  </aug>
  <source>Biometrics</source>
  <publisher>JSTOR</publisher>
  <pubdate>1986</pubdate>
  <fpage>121</fpage>
  <lpage>-130</lpage>
</bibl>

<bibl id="B12">
  <title><p>Regression modeling strategies: with applications to linear models,
  logistic and ordinal regression, and survival analysis</p></title>
  <aug>
    <au><snm>Harrell Jr</snm><fnm>FE</fnm></au>
  </aug>
  <publisher>Springer</publisher>
  <pubdate>2015</pubdate>
</bibl>

<bibl id="B13">
  <title><p>Introduction to robust estimation and hypothesis
  testing</p></title>
  <aug>
    <au><snm>Wilcox</snm><fnm>RR</fnm></au>
  </aug>
  <publisher>Academic press</publisher>
  <pubdate>2011</pubdate>
</bibl>

<bibl id="B14">
  <title><p>Testing statistical hypotheses</p></title>
  <aug>
    <au><snm>Lehmann</snm><fnm>EL</fnm></au>
    <au><snm>Romano</snm><fnm>JP</fnm></au>
  </aug>
  <publisher>Springer Science \& Business Media</publisher>
  <pubdate>2006</pubdate>
</bibl>

<bibl id="B15">
  <title><p>Controlling the false discovery rate: a practical and powerful
  approach to multiple testing</p></title>
  <aug>
    <au><snm>Benjamini</snm><fnm>Y</fnm></au>
    <au><snm>Hochberg</snm><fnm>Y</fnm></au>
  </aug>
  <source>Journal of the Royal statistical society: series B
  (Methodological)</source>
  <publisher>Wiley Online Library</publisher>
  <pubdate>1995</pubdate>
  <volume>57</volume>
  <issue>1</issue>
  <fpage>289</fpage>
  <lpage>-300</lpage>
</bibl>

<bibl id="B16">
  <title><p>Controlling the false discovery rate in behavior genetics
  research</p></title>
  <aug>
    <au><snm>Benjamini</snm><fnm>Y</fnm></au>
    <au><snm>Drai</snm><fnm>D</fnm></au>
    <au><snm>Elmer</snm><fnm>G</fnm></au>
    <au><snm>Kafkafi</snm><fnm>N</fnm></au>
    <au><snm>Golani</snm><fnm>I</fnm></au>
  </aug>
  <source>Behavioural brain research</source>
  <publisher>Elsevier</publisher>
  <pubdate>2001</pubdate>
  <volume>125</volume>
  <issue>1-2</issue>
  <fpage>279</fpage>
  <lpage>-284</lpage>
</bibl>

<bibl id="B17">
  <title><p>The control of the false discovery rate in multiple testing under
  dependency</p></title>
  <aug>
    <au><snm>Benjamini</snm><fnm>Y</fnm></au>
    <au><snm>Yekutieli</snm><fnm>D</fnm></au>
  </aug>
  <source>Annals of statistics</source>
  <publisher>JSTOR</publisher>
  <pubdate>2001</pubdate>
  <fpage>1165</fpage>
  <lpage>-1188</lpage>
</bibl>

<bibl id="B18">
  <title><p>Statistical significance for genomewide studies</p></title>
  <aug>
    <au><snm>Storey</snm><fnm>JD</fnm></au>
    <au><snm>Tibshirani</snm><fnm>R</fnm></au>
  </aug>
  <source>Proceedings of the National Academy of Sciences</source>
  <publisher>National Acad Sciences</publisher>
  <pubdate>2003</pubdate>
  <volume>100</volume>
  <issue>16</issue>
  <fpage>9440</fpage>
  <lpage>-9445</lpage>
</bibl>

</refgrp>
} 




\begin{table}[h!]
	\centering
	\scriptsize
	\caption{Detected `outlier' audiologists from AAA of CHEARS. Each audiologist's coefficient estimate is compared with the 10\% truncated mean of all audiologists' coefficient estimates.}
	\begin{tabular}{ccccc}
		\hline \hline
		Alternative	&	Power		&  $\widehat{FDR}$  & Outlier Audiologists& Adujsted Outlier Audiologists \\ \hline
		\multirow{3}{*}{$H_{1,j}: \Big|\boldsymbol{L}_{10\%, j}\boldsymbol{\beta}\Big|=5$}&\multirow{3}{*}{0.80}&\multirow{3}{*}{0.72}&2, 4, 8, 13, 14, 15, 16, 17, &\multirow{3}{*}{4, 13, 16, 41, 48}\\
		&&& 22, 24, 28, 36, 39, 41, 42, 47, & \\
		&&& 48, 49, 52, 54, 55, 57, 58, 59&\\\hline
		\multirow{2}{*}{$H_{1,j}:\Big|\boldsymbol{L}_{10\%, j}\boldsymbol{\beta}\Big|=5$}&\multirow{2}{*}{0.47}&\multirow{2}{*}{0.50}&4, 13, 14, 15, &\multirow{2}{*}{4, 13, 48}\\
		&&&22, 48, 54, 55 & \\\hline
		$H_{1,j}: \Big|\boldsymbol{L}_{10\%, j}\boldsymbol{\beta}\Big|=10$&0.80&0.44&4, 13, 14, 55 &4, 13 \\\hline
		$H_{1,j}: \Big|\boldsymbol{L}_{10\%, j}\boldsymbol{\beta}\Big|=10$&0.86&0.50&4, 13, 14, 22, 55&4, 13 \\\hline
		\multirow{2}{*}{-}&\multirow{2}{*}{(0.55, 1.00)}&\multirow{2}{*}{0.28}&4, 13, 15, 16, 17, 22, &4, 13, 16, 22,\\
		&&&24, 40, 41, 45, 48, 63&24, 40, 41, 48, 63\\\hline
	\end{tabular} \label{real.gee.1}
	\begin{tablenotes}
		\item[*] The last row reports the results from using $\alpha=0.05$ as the significance level for rejecting tests $H_{0,1},\ldots, H_{0,68}$. The range of the power of these 68 hypothesis tests under $\alpha=0.05$ was reported in the Power column.
	\end{tablenotes}
\end{table}

\begin{table}[h!]
	\centering
	\scriptsize
	\caption{Detected `outlier' audiologists from AAA of CHEARS. Each audiologist's coefficient estimate is compared with untruncated mean of all audiologists' coefficient estimates.}
	\begin{tabular}{ccccc}
		\hline \hline
		Alternative	&	Power		&  $\widehat{FDR}$  & Outlier Audiologists& Adujsted Outlier Audiologists \\ \hline
		\multirow{3}{*}{$H_{1,j}: \Big|\boldsymbol{L}_j\boldsymbol{\beta}\Big|=5$}&\multirow{3}{*}{0.80}&\multirow{3}{*}{0.72}&2, 4, 8, 13, 14, 15, 16, 17, &\multirow{3}{*}{4, 13, 16, 41, 48}\\
		&&& 22, 24, 28, 36, 39, 41, 42, 47, & \\
		&&& 48, 49, 52, 54, 55, 57, 58, 59&\\\hline
		\multirow{2}{*}{$H_{1,j}:\Big|\boldsymbol{L}_j\boldsymbol{\beta}\Big|=5$}&\multirow{2}{*}{0.47}&\multirow{2}{*}{0.50}&4, 13, 14, 15, &\multirow{2}{*}{4, 13, 22, 48} \\
		&&&22, 48, 54, 55, 59&\\\hline
		$H_{1,j}: \Big|\boldsymbol{L}_j\boldsymbol{\beta}\Big|=10$&0.80&0.44&4, 13, 14, 55 &4, 13 \\\hline
		$H_{1,j}: \Big|\boldsymbol{L}_j\boldsymbol{\beta}\Big|=10$&0.85&0.50&4, 13, 14, 22, 55&4, 13 \\\hline
		\multirow{2}{*}{-}&\multirow{2}{*}{(0.55, 1.00)}&\multirow{2}{*}{0.28}&4, 13, 15, 16, 17, 22, &4, 13, 16, 17, \\
		&&&24, 40, 41, 45, 48, 63&22, 24, 40, 41, 48\\
		\hline\hline
	\end{tabular} \label{real.gee.2}
	\begin{tablenotes}
		\item[*] The last row reports the results from using $\alpha=0.05$ as the significance level for rejecting tests $H_{0,1},\ldots, H_{0,68}$. The range of the power of these 68 hypothesis tests under $\alpha=0.05$ was reported in the Power column.
	\end{tablenotes}
\end{table}

\begin{figure}[htb]
	\centering
	\includegraphics[width=5in]{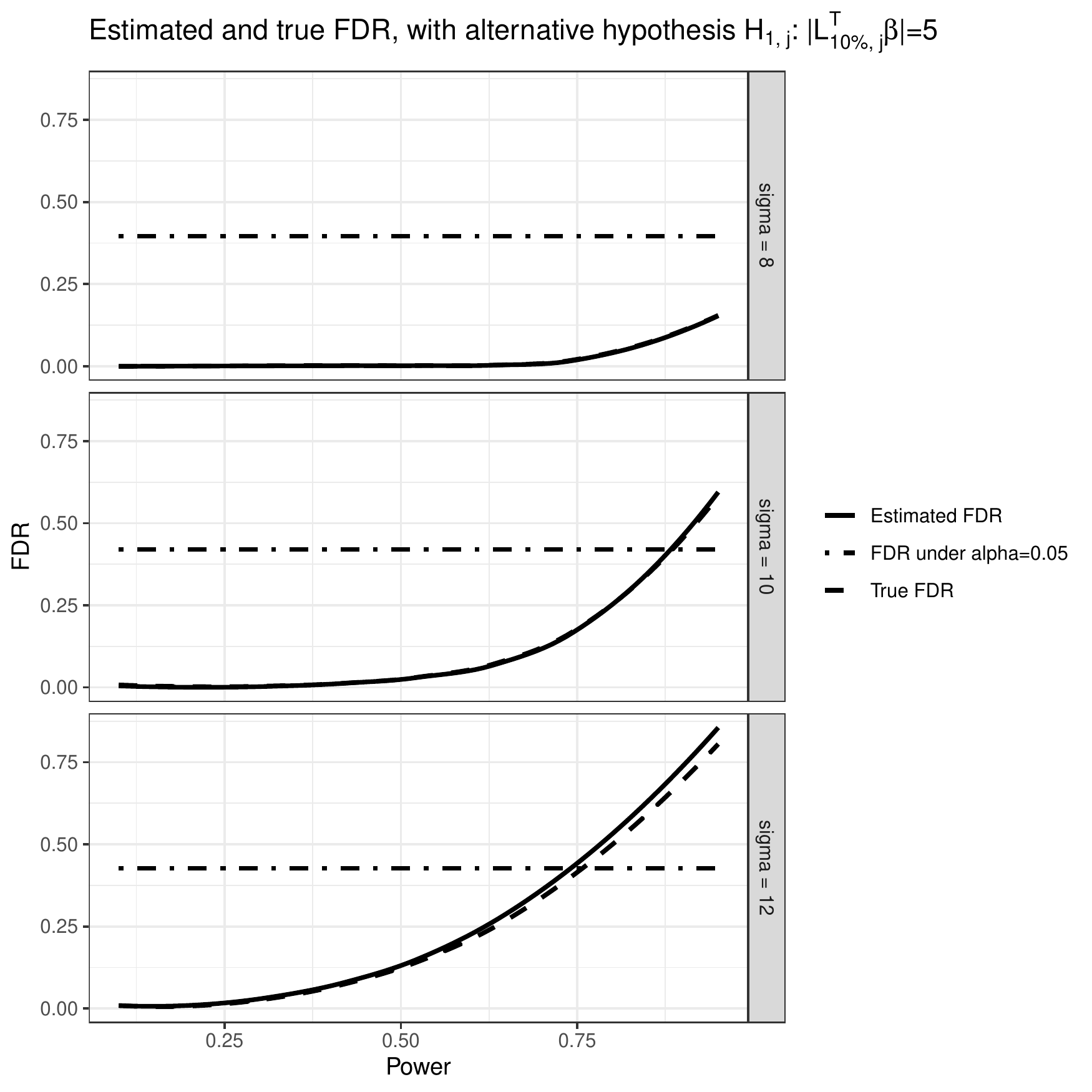}
	\caption{FDR vs. Power decision plot for single measurement simulation. The alternative hypothesis is $H_{1,j}: \Big|\boldsymbol{L}^T_{10\%, j}\boldsymbol{\beta}\Big|=5$. The solid curve is the estimated FDR based on Equation (\ref{final_fdr_est}) averaged over 300 simulation replicates, and the dashed curve is the empirical true FDR calculated by averaging the proportions of false discoveries $\frac{\boldsymbol{V}(\phi)}{\boldsymbol{R}(\phi)}$ over 300 simulation replicates. The black horizontal dot-dash line represents the empirical true FDR calculated by averaging the proportions of false discoveries over 300 simulation replicates when using $\alpha=0.05$ as the significance level. The solid and dashed curves are overlapped on the top panel.}
	\label{one.ear.fdr}
\end{figure}

\begin{figure}[htb]
	\begin{subfigure}{.4\textwidth}
		\centering
		\includegraphics[width=2.4in]{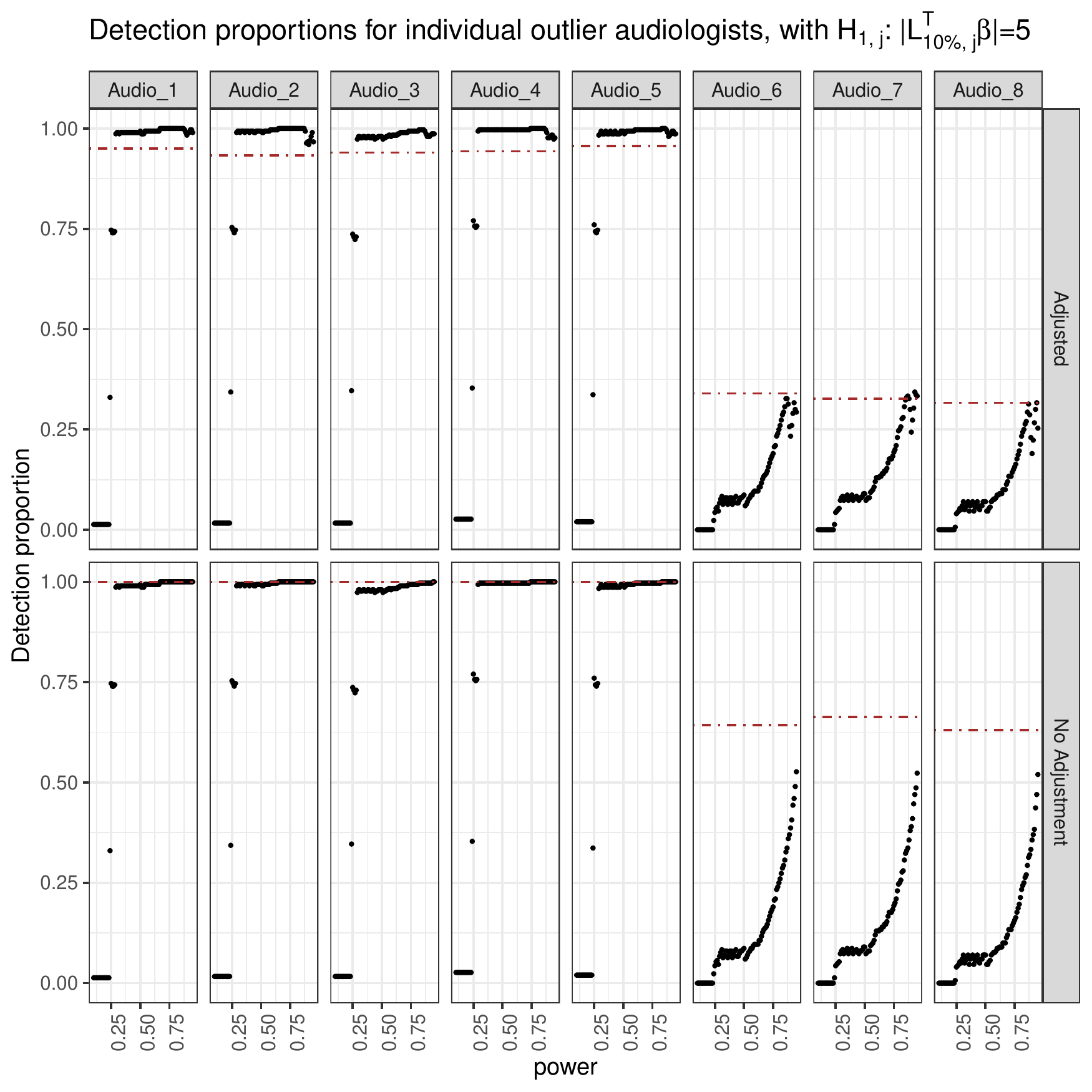}  
		\caption{True positive proportions for Audiologists 1 - 8.}
		\label{fig2.a}
	\end{subfigure}
	\begin{subfigure}{.4\textwidth}
		\centering
		\includegraphics[width=2.4in]{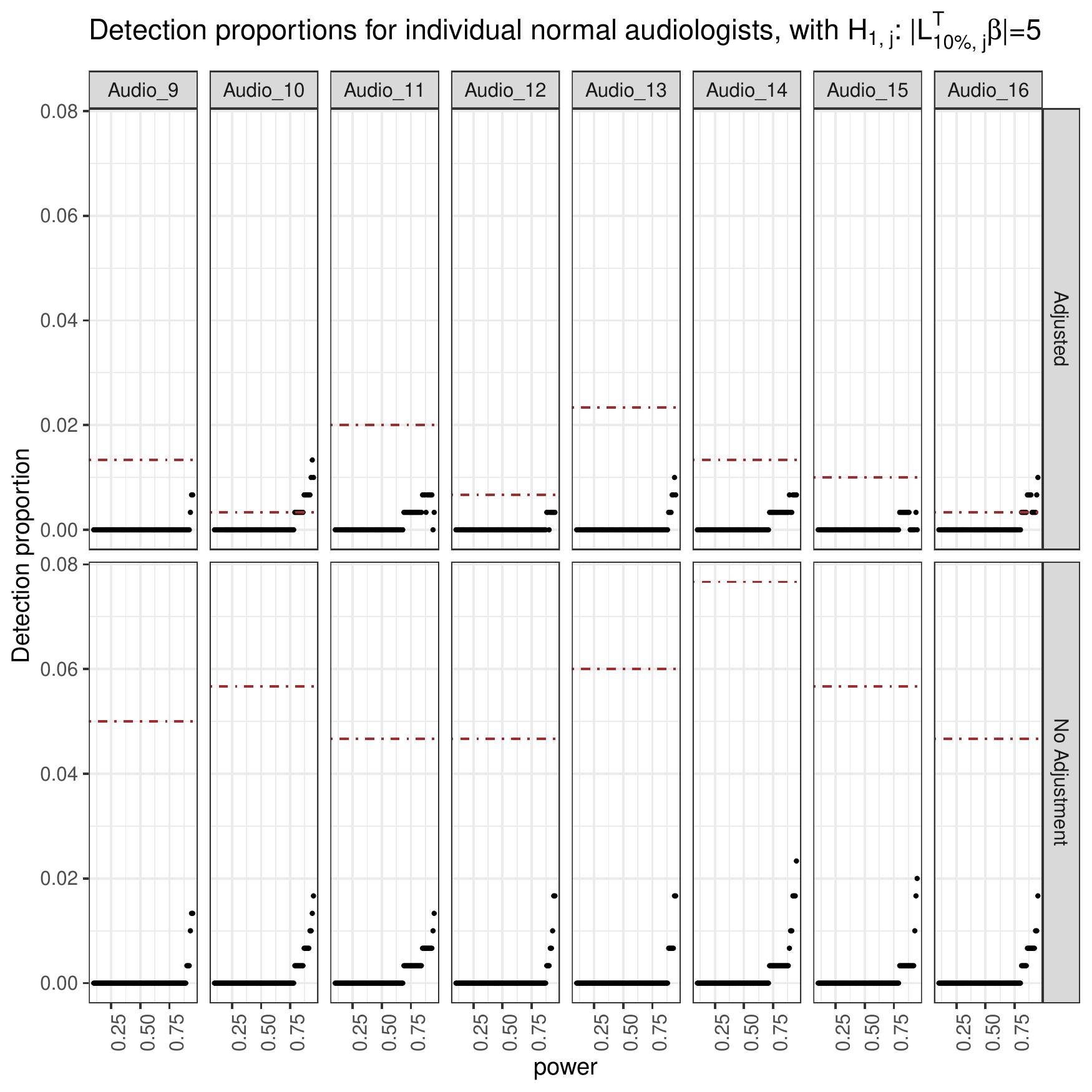}  
		\caption{ False positive proportions for Audiologists 9 - 16.}
		\label{fig2.b}
	\end{subfigure}
	\caption{This figure shows the true positive proportions for the true `outlier' audiologists and false positive proportions for the true `normal' audiologists for single measurement simulation with $\sigma=8$. The top panel in each subfigure is the result by performing the FDR-based adjustment, while the bottom panel in each subfigure is the result without FDR-based adjustment. The horizontal dot-dash line represents the corresponding true or false positive proportion for each audiologist if we use $\alpha=0.05$ as the significance level for rejecting the null hypotheses.
	}
	\label{one.ear.detect.sigma8}
\end{figure}

\begin{figure}[htb]
		\begin{subfigure}{.4\textwidth}
		\centering
		\includegraphics[width=2.4in]{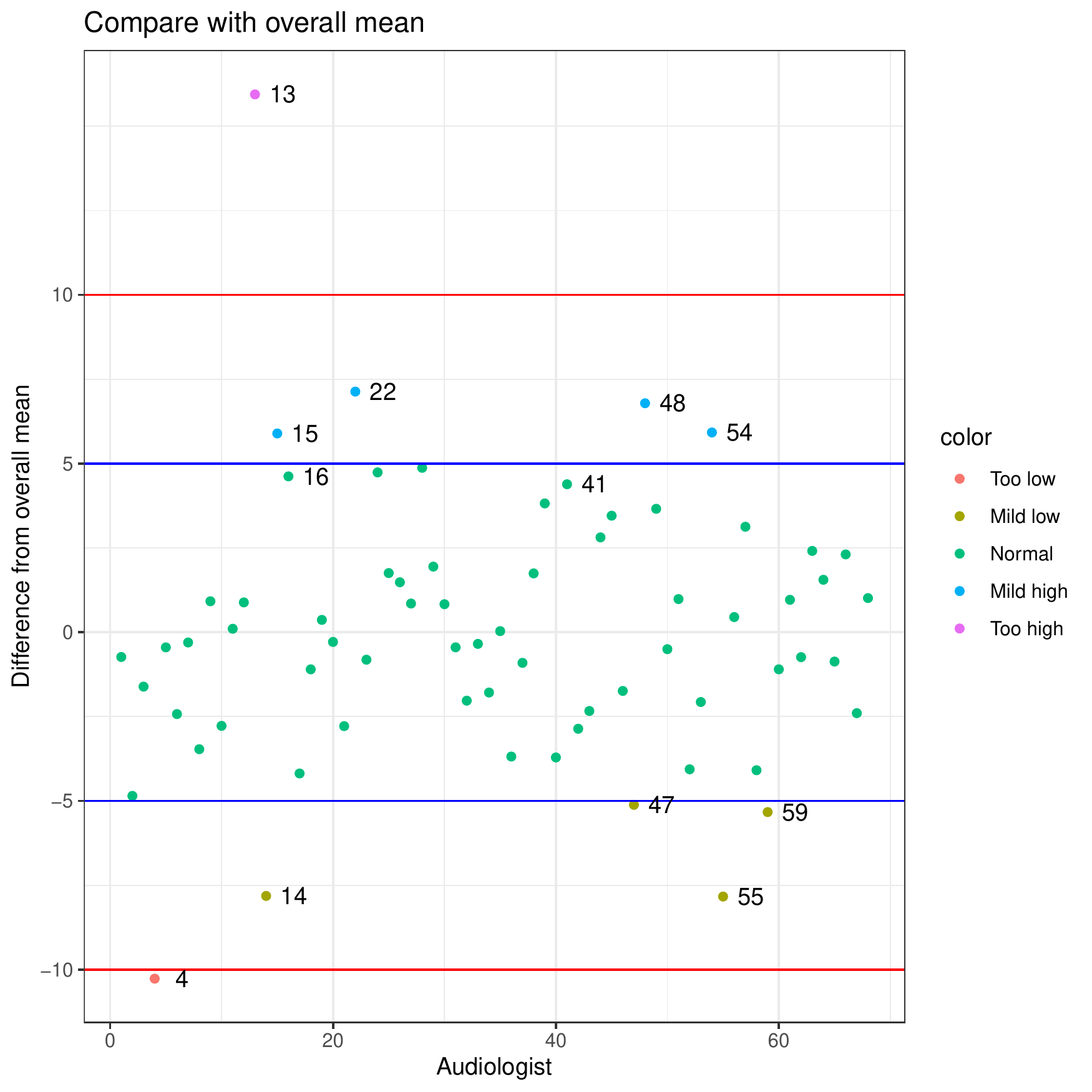}  
		\caption{}
		\label{fig3.a}
	\end{subfigure}
	\begin{subfigure}{.4\textwidth}
		\centering
		\includegraphics[width=2.4in]{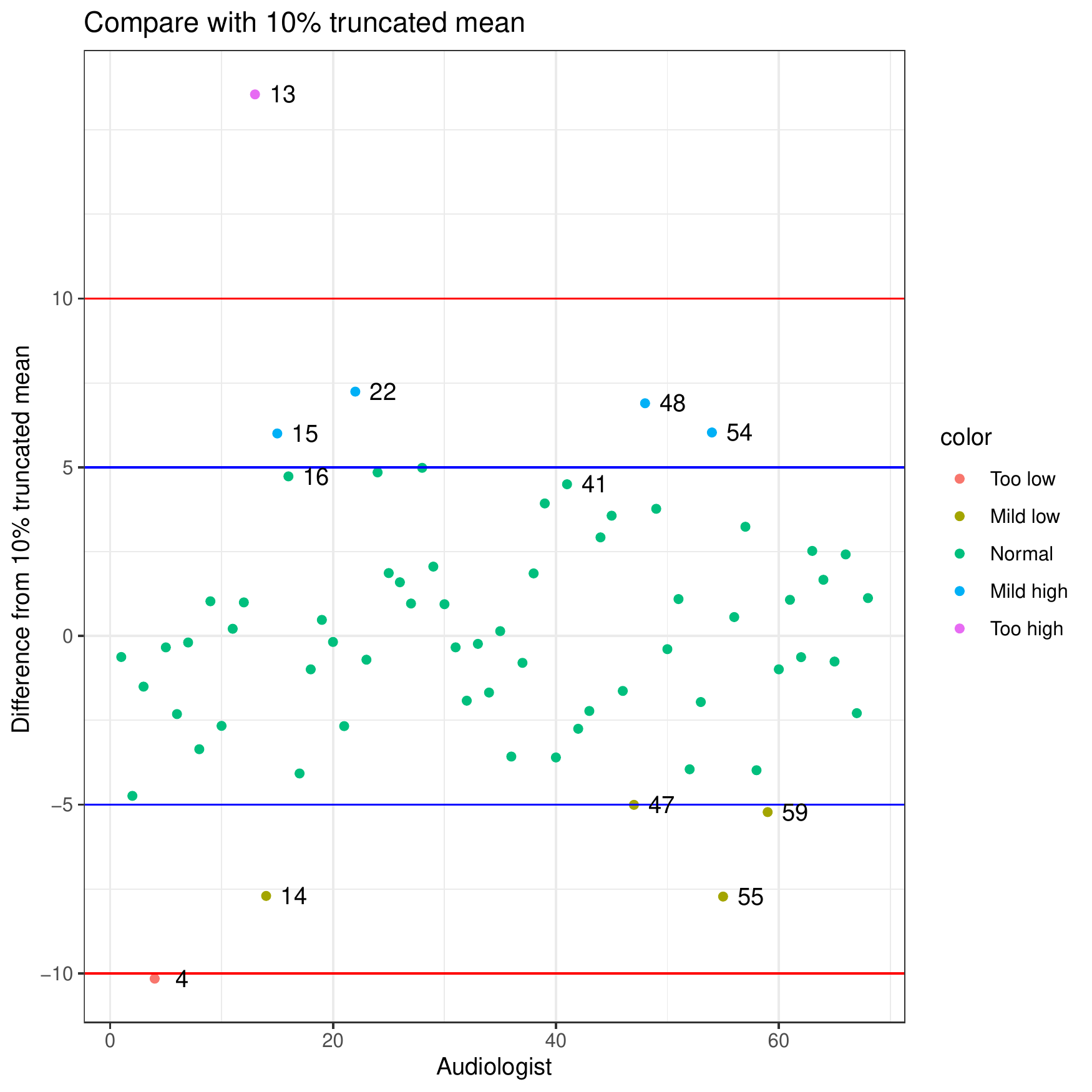}  
			\caption{}
		\label{fig3.b}
	\end{subfigure}
	\caption{ (a) Subtracting each audiologist's coefficient estimate by the untruncated mean of all audiologists' coefficient estimates,; (b) Subtracting each audiologist's coefficient estimate by the 10\% truncated mean of all audiologists' coefficient estimates. }
	\label{gee_ear_plot_mean}
\end{figure}

\begin{figure}[htb]
	\begin{subfigure}{.4\textwidth}
		\centering
		\includegraphics[width=2in]{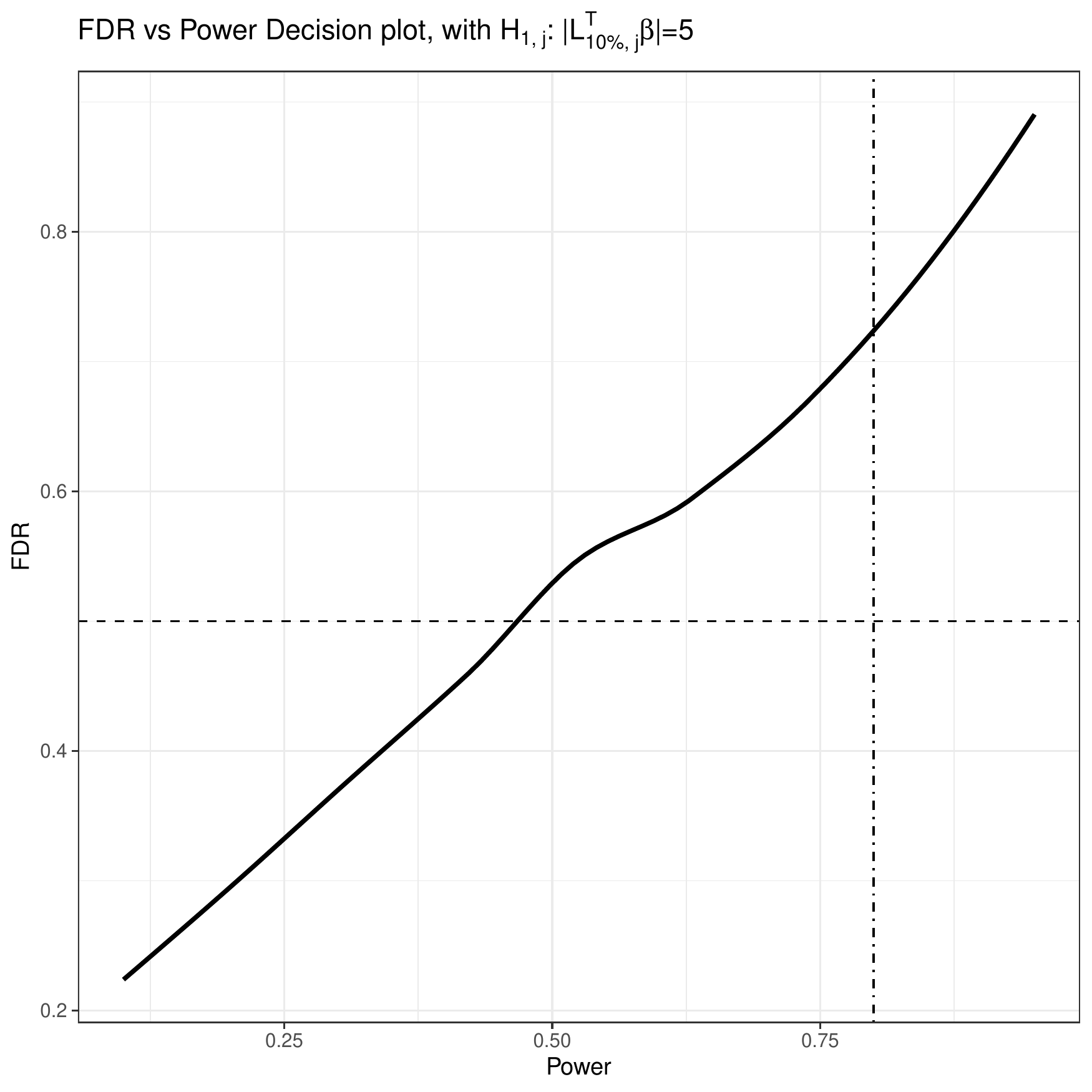}  
		\caption{}
		\label{fig4.a}
	\end{subfigure}
\hfill
	\begin{subfigure}{.4\textwidth}
		\centering
		\includegraphics[width=2in]{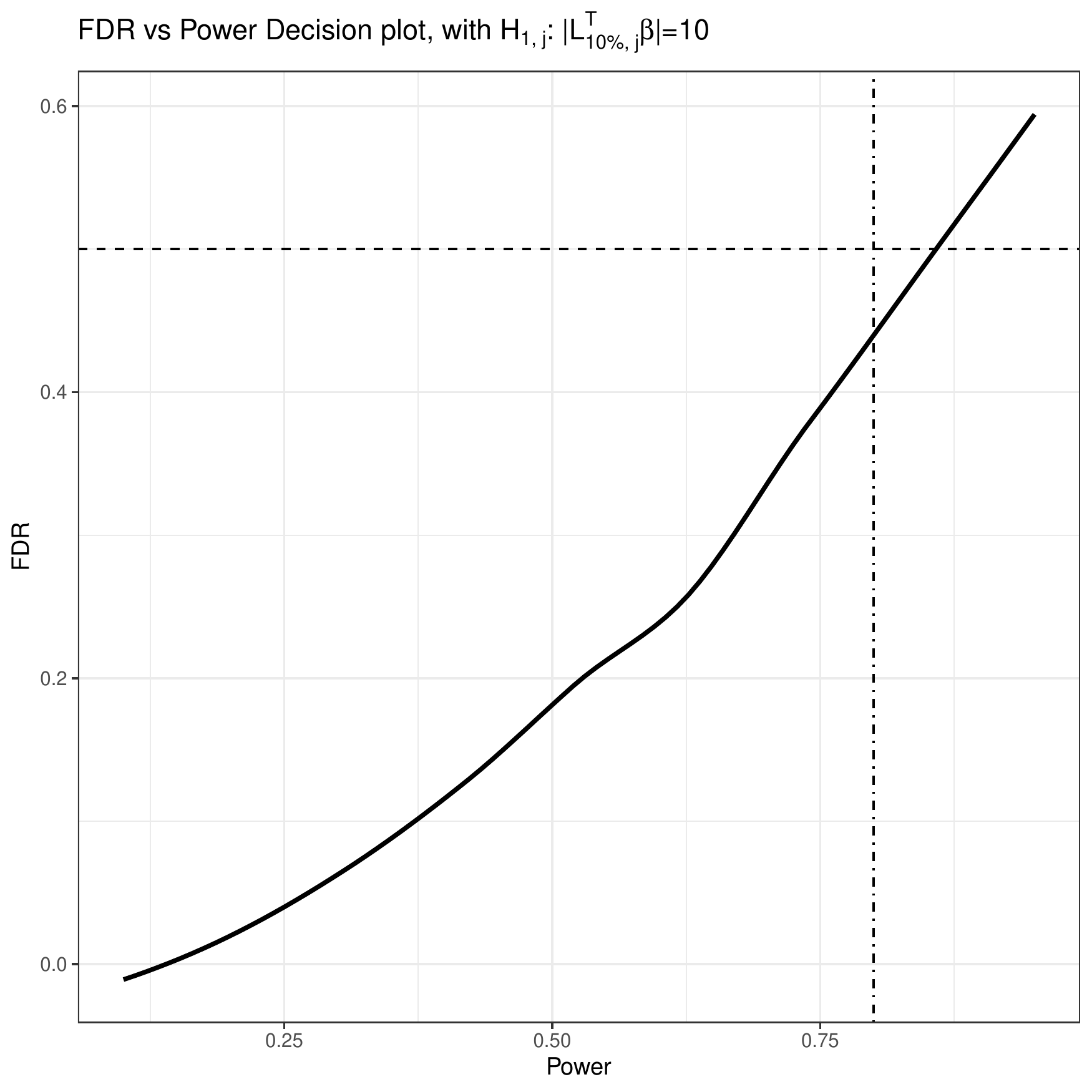}  
		\caption{}
		\label{fig4.b}
	\end{subfigure}
\hfill
	\begin{subfigure}{.4\textwidth}
		\centering
		\includegraphics[width=2in]{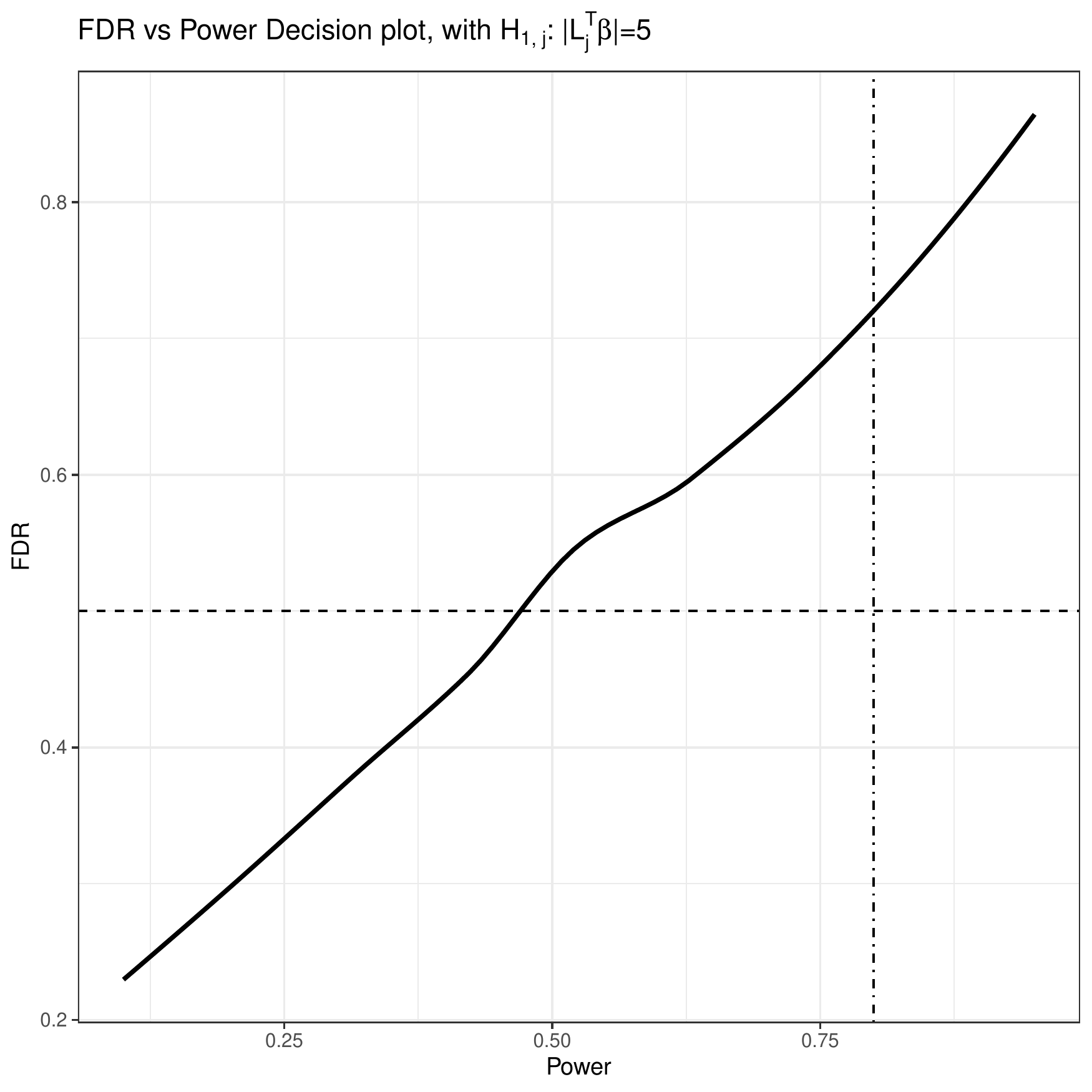}  
		\caption{}
		\label{fig4.c}
	\end{subfigure}
\hfill
	\begin{subfigure}{.4\textwidth}
		\centering
		\includegraphics[width=2in]{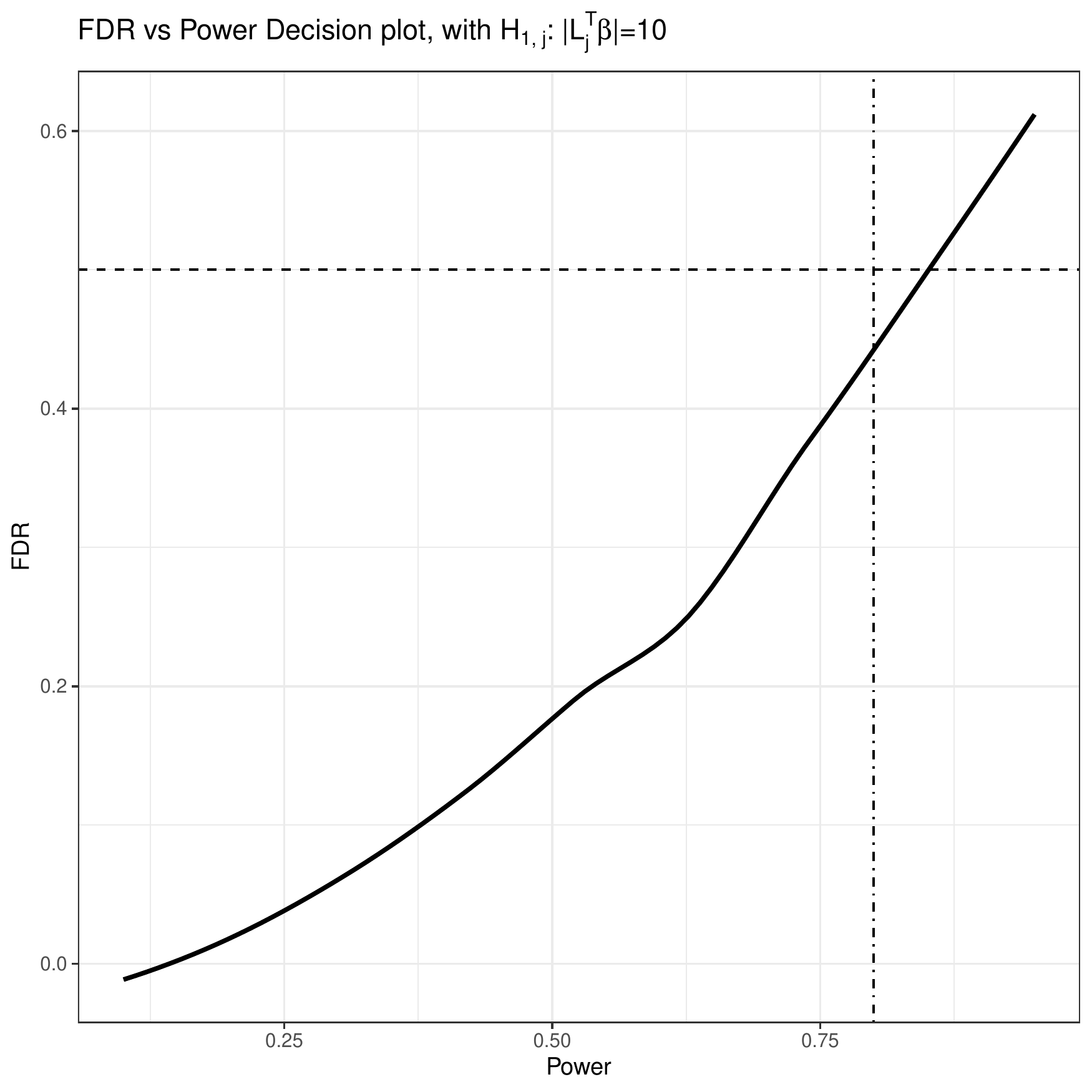}  
		\caption{}
		\label{fig4.d}
	\end{subfigure}

	\caption{FDR vs. Power decision plot for detecting `outlier' audiologists, where (a): $H_{1,j}: \Big|\boldsymbol{L}^T_{10\%, j}\boldsymbol{\beta}\Big|=5$; (b): $H_{1,j}: \Big|\boldsymbol{L}^T_{10\%, j}\boldsymbol{\beta}\Big|=10$; (c): $H_{1,j}: \Big|\boldsymbol{L}^T_j\boldsymbol{\beta}\Big|=5$; and (d): $H_{1,j}: \Big|\boldsymbol{L}^T_j\boldsymbol{\beta}\Big|=10$.
		The dot-dash and dashed lines are produced by fixing power at 0.8 or the FDR at 0.5, respectively.}
	\label{gee_ear_fdrvspower}
\end{figure}


\end{backmatter}
\end{document}